\documentclass[a4paper,11pt]{article}
\pdfoutput=1 
\usepackage[UKenglish]{babel}
\usepackage{microtype,xspace}
\usepackage[T2A]{fontenc}
\usepackage[utf8]{inputenc}	

\usepackage{jheppub} 
\usepackage{bm,amssymb,amsmath,amsfonts,slashed,graphicx,multirow,soul,mathtools,xspace,array,comment,color}
\hypersetup{colorlinks=true, allcolors=blue}  
\usepackage{adjustbox}
\usepackage{siunitx}
\usepackage{bm} 
\usepackage{float}
\usepackage{placeins}
\allowdisplaybreaks
\usepackage{ bbold }
\usepackage[normalem]{ulem} 
\usepackage{enumerate}
\usepackage{multirow}
\usepackage{booktabs}
\usepackage{xparse}
\usepackage{etoolbox}
\usepackage{physics}
\usepackage{hypcap}
\usepackage{siunitx}
\usepackage{xcolor}
\usepackage{afterpage}
\usepackage{comment}
\usepackage[compat=1.1.0]{tikz-feynman}
\usepackage[capitalise]{cleveref}
\usepackage{caption}
\usepackage{subcaption}
\usepackage{scalefnt,pstricks}
\usepackage{listings}

\def\be{\begin{equation}}
\def\ee{\end{equation}}
\newcommand{\TeV}{\text{TeV}}
\newcommand{\GeV}{\text{GeV}}

\newcommand{\ol}{\textsc{OpenLoops}}

\arxivnumber{} 

\providecommand{\href}[2]{#2}

\newcommand{\loi}{LO\text{-}I}
\newcommand{\nloi}{NLO\text{-}I}
\newcommand{\borni}{Born\text{-}I}
\newcommand{\FTa}[1]{aFT\text{-}#1}

\newcommand{\mwcom}[1]{}

\newcommand{\pt}{{p_{\text{\scalefont{0.77}T}}}}

\newcommand{\noun}[1]{{\scshape #1}}

\newcommand{\POWHEG}{\noun{Powheg}}

\newcommand{\POWHEGBOX}{\noun{Powheg-Box}}

\newcommand{\minnlo}{{\noun{MiNNLO$_{\rm PS}$}}\xspace}

\newcommand{\citere}[1]{ref.\,\cite{#1}}

\newcommand{\citeres}[1]{refs.\,\cite{#1}}

\newcommand{\eqn}[1]{eq.\,(\ref{#1})}

\newcommand{\fig}[1]{figure\,\ref{#1}}

\newcommand{\tab}[1]{table\,\ref{#1}}
\newcommand{\sct}[1]{section~\ref{#1}}

\newcommand{\app}[1]{appendix~\ref{#1}}

\usepackage{etoolbox}
\makeatletter
\patchcmd{\@sect}{#8}{\boldmath #8}{}{}
\let\ori@chapter\@chapter
\def\@chapter[#1]#2{\ori@chapter[\boldmath#1]{\boldmath#2}}
\makeatother

\title{\boldmath NNLO+PS Higgs-pair production in MiNNLO$_{\textrm{PS}}$}

\emailAdd{garosi@mpp.mpg.de}
\emailAdd{marius.wiesemann@mpp.mpg.de}
\emailAdd{zanderi@mpp.mpg.de}

\author[a]{Francesco Garosi,}
\author[a]{Marius Wiesemann,}
\author[a,b]{Giulia Zanderighi}

\affiliation[a]{Max-Planck-Institut f\"ur Physik, Boltzmannstra\ss e 8, 85748 Garching, Germany}
\affiliation[b]{Physik Department T31, Technische Universität München,
James-Franck-Straße 1, D-85748 Garching, Germany}

\abstract{
We consider Higgs-boson pair production in gluon fusion at hadron colliders and match next-to-next-to-leading-order (NNLO) QCD corrections to parton showers within the \minnlo{} framework. Since the full top-quark mass dependence at this order is not available, finite top-quark mass effects are incorporated through approximations based on the exact NLO QCD result, using the available two-loop amplitude in the full theory. Specifically, the Born, single-virtual, single-real and double-real contributions are included exactly, while the real--virtual and double-virtual corrections are approximated. We consider different approximations for the latter to assess the associated uncertainties. We validate our predictions against fixed-order NNLO QCD results and compare with existing NNLO calculations matched to parton shower from \textsc{Geneva}, where in some cases we find noticeable differences. Finally, we present phenomenological results for different Higgs-decay channels and variations of the trilinear Higgs coupling.

\vspace{0.2cm}\noindent Our \minnlo{} generator for Higgs-boson pair production is available within the \textsc{Powheg-Box-Res} framework.
}


\preprint{MPP-2026-81}

\begin{document}
\maketitle
\flushbottom


\section{Introduction}
\label{sec:intro}

The exploration of the electroweak symmetry breaking mechanism remains a central goal of the physics programme at the Large Hadron Collider (LHC). Following the discovery of the Higgs boson in 2012 \cite{Aad:2012tfa,Chatrchyan:2012xdj}, precision measurements of its properties provide stringent tests of the Standard Model (SM) and sensitivity to possible effects of physics beyond it. While many of the Higgs couplings have already been measured with remarkable precision, the structure of the Higgs potential is still largely unconstrained. In this context, Higgs-boson pair production plays a special role, as it provides direct access to the Higgs self-coupling.
 
 At the LHC, Higgs-boson pair ($HH$) production is dominated by gluon fusion, where the coupling of the Higgs bosons to gluons is mediated by heavy-quark loops. Owing to its loop-induced nature, this process is particularly challenging from a theoretical point of view. In fact, the leading-order contribution already involves one-loop amplitudes, such that higher-order QCD corrections require the computation of multi-loop scattering amplitudes with several kinematic scales.

A commonly employed simplification is given by the heavy-top limit, in which the top quark is integrated out and the gluon--Higgs interaction is described by an effective operator. In this approximation, the loop-induced leading-order process is replaced by a tree-level interaction, thereby effectively lowering the loop order of the corresponding perturbative corrections by one. While this approach has proven to be highly successful for single-Higgs production, where typical energy scales are set by the Higgs mass, its validity is much more limited for Higgs-boson pair production. In this case, the relevant kinematic scales are significantly larger, starting at the $HH$ threshold of twice the Higgs-boson mass and extending to even higher invariant masses. As a consequence, finite top-quark mass effects play a crucial role, and their inclusion is mandatory for reliable predictions.

At the same time, the computation of higher-order corrections with full mass dependence is considerably more involved than in single-Higgs production. Higgs-boson pair production constitutes a genuine $2 \to 2$ scattering process, leading to multi-scale loop integrals with non-trivial kinematic dependence already at the amplitude level. This renders the calculation of higher-order corrections in the full theory significantly more challenging than for single-Higgs production, where NNLO QCD corrections are known including full top-quark mass effects~\cite{Czakon:2021yub,Niggetiedt:2024nmp}.

The leading-order (LO) predictions for $HH$ production were obtained long ago~\cite{Glover:1987nx,Plehn:1996wb}, and next-to-leading-order (NLO) QCD corrections have been computed both in the heavy-top approximation and including full top-quark mass effects~\cite{Dawson:1998py,Borowka:2016ehy,Borowka:2016ypz}. Next-to-next-to-leading-order (NNLO) QCD corrections are available in the heavy-top limit, both for the inclusive cross section~\cite{deFlorian:2013jea,Grigo:2014jma} and at the fully differential level~\cite{deFlorian:2016uhr}. These predictions have also been combined with soft-gluon resummation~\cite{deFlorian:2015moa,Shao:2013bz}, and inclusive N$^3$LO corrections have recently been obtained in the same approximation~\cite{Davies:2018ood,Chen:2021ibm}. In addition, the fully differential NNLO predictions in the heavy-top limit have been improved through the inclusion of top-quark mass effects at NLO, leading to the full theory approximated (FT-approx) result \cite{Grazzini:2018bsd}. Despite this significant progress, the inclusion of full top-quark mass effects in fully differential predictions beyond NLO remains one of the main open challenges for precision studies of Higgs-boson pair production.

For realistic phenomenological applications, it is essential to combine higher-order QCD calculations with parton-shower simulations, enabling fully exclusive event generation. While NLO+PS predictions for Higgs-boson pair production have been available for some time~\cite{Frederix:2014hta}, NNLO+PS calculations are considerably more involved. First NNLO+PS results for this process have been obtained within the \textsc{Geneva} framework, both in the heavy-top approximation and including full top-quark mass effects~\cite{Alioli:2022dkj,Alioli:2025xcu}. The construction of such predictions remains highly non-trivial, in particular in the presence of full mass effects. A comparison of these results with a completely independent NNLO+PS implementation is therefore particularly valuable, since it allows one to assess whether the results agree within the quoted uncertainties.

In this work, we present NNLO QCD predictions for Higgs-boson pair production matched to parton showers, including top-quark mass effects. Our calculation is performed within the \textsc{MiNNLO$_{\rm PS}$} framework~\cite{Monni:2019whf}. Finite top-quark mass effects are incorporated by exploiting their exact treatment at NLO in QCD, which is consistently embedded into the NNLO+PS calculation. To this end, we consider different strategies to propagate mass effects beyond NLO accuracy, allowing us to assess the associated theoretical uncertainties.
Furthermore, we make use of available two-loop information in the full theory, including approximate results based on different computational approaches, some of which are particularly efficient. This enables the construction of a numerically stable and computationally efficient Monte Carlo implementation. The resulting event generator provides NNLO+PS accurate predictions including top-quark mass effects, supports arbitrary values of the trilinear Higgs coupling, and therefore is well suited for precision phenomenological studies and fits to data. The code is made publicly available together with this publication.

The paper is organised as follows.
In \sct{sec:outline}, we briefly review the \textsc{MiNNLO$_{\rm PS}$}
framework and describe the implementation for Higgs-boson pair production.
Section\,\ref{sec:topmass} introduces the different approximations used to
account for top-quark mass effects.
Phenomenological results for stable Higgs bosons are presented in
\sct{sec:results}, including comparisons to existing calculations.
Results with Higgs decays and with modified trilinear Higgs coupling are
discussed in sections\,\ref{sec:higgsdecays} and \ref{sec:trilinear},
respectively.
Finally, we summarise our findings and discuss future directions in
\sct{sec:conclusions}.
Appendix~\ref{app:H1_H2_ggHH} collects the expressions for the hard function for $HH$ production.

\section{Outline of the calculation}
\label{sec:outline}
We consider the process

\be
pp\to HH+X \label{eq:process}\,,
\ee
inclusive over the radiation of extra particles $X$, and we include NNLO corrections in QCD perturbation theory.
 
\begin{figure}[t]
    \centering

    \begin{subfigure}[b]{0.3\textwidth}
        \centering
        \includegraphics[width=\linewidth]{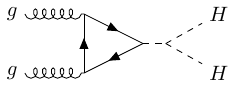}
        \caption{LO: triangle}
        \label{fig:triangle_lo}
    \end{subfigure}
    \hspace{0.05\textwidth}
    \begin{subfigure}[b]{0.3\textwidth}
        \centering
        \includegraphics[width=\linewidth]{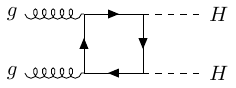}
        \caption{LO: box}
        \label{fig:box_lo}
    \end{subfigure}

    \vspace{0.3cm}

    \begin{subfigure}[b]{0.3\textwidth}
        \centering
        \includegraphics[width=\linewidth]{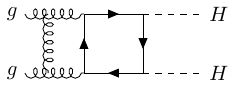}
        \caption{NLO: virtual}
        \label{fig:2loops}
    \end{subfigure}
    \hspace{0.05\textwidth}
    \begin{subfigure}[b]{0.3\textwidth}
        \centering
        \includegraphics[width=\linewidth]{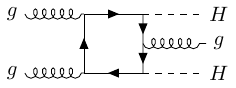}
        \caption{NLO: real}
        \label{fig:gghhg}
    \end{subfigure}

    \vspace{0.3cm}

    \begin{subfigure}[b]{0.3\textwidth}
        \centering
        \includegraphics[width=\linewidth]{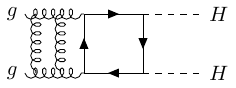}
        \caption{NNLO: double virtual}
        \label{fig:3loops}
    \end{subfigure}\hfill
    \begin{subfigure}[b]{0.3\textwidth}
        \centering
        \includegraphics[width=\linewidth]{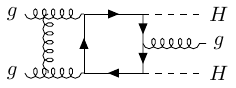}
        \caption{NNLO: real--virtual}
        \label{fig:gghhg2l}
    \end{subfigure}\hfill
    \begin{subfigure}[b]{0.3\textwidth}
        \centering
        \includegraphics[width=\linewidth]{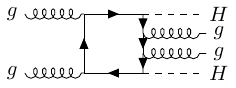}
        \caption{NNLO: double real}
        \label{fig:gghhgg}
    \end{subfigure}

    \caption{Sample diagrams for $pp\to HH$ up to NNLO in QCD.}
    \label{fig:feyn_diagrams}
\end{figure}

The LO process proceeds via gluon fusion and is loop induced.
Example Feynman diagrams are shown in \fig{fig:feyn_diagrams}\,(a) and (b),
corresponding to the triangle contribution, which is sensitive to the trilinear Higgs coupling,
and the box contribution, respectively. We include only top-quark loops, which provide by far
the dominant contribution due to the large Yukawa coupling. At NLO, owing to the loop-induced
nature of the process, the virtual contribution is given by the two-loop $gg\to HH$ amplitude,
see \fig{fig:feyn_diagrams}\,(c), while the real emission process $pp\to HHj$ is also
loop induced, see \fig{fig:feyn_diagrams}\,(d). At NNLO, the double-virtual $gg\to HH$,
the real--virtual $pp\to HHj$ and the double-real $pp\to HHjj$ contributions arise from the
corresponding three-loop, two-loop and one-loop amplitudes, respectively, with representative
diagrams shown in \fig{fig:feyn_diagrams}\,(e)--(g). While the LO process is purely
gluon initiated, additional partonic channels involving also quarks and antiquarks contribute
from NLO onwards.

In the full theory, both the three-loop $gg\to HH$ and the two-loop $pp\to HHj$
amplitudes are not yet known. An alternative approach is provided by the
approximation of an infinitely heavy top-quark mass, commonly referred to as
the heavy-top limit (HTL). In this framework, the top quark is integrated out and
the LO contribution reduces to a tree-level amplitude induced by effective
Higgs--gluon contact interactions. As a result, higher-order corrections involve
amplitudes with one loop fewer compared to the full theory. While the HTL does
not provide an accurate description of the full $HH$ process, due to the limited
validity of the infinite top-mass approximation, it can be used to construct
approximations for the missing amplitudes. This will be discussed in detail in
\sct{sec:topmass}.

We combine NNLO QCD corrections to $HH$ production with a parton-shower simulation
by means of the \minnlo{} method. Our $HH$ \minnlo{} generator is implemented within
the \textsc{Powheg-Box-Res} framework~\cite{Jezo:2015aia}. Originally developed for
colour-singlet (F) production and first applied to single-boson processes
\cite{Monni:2019whf, Monni:2020nks}, we employ here the extension of \minnlo{} to
multi-boson reactions \cite{Lombardi:2020wju}. In its most recent developments, the \minnlo{}
approach has also been extended to heavy-quark pair production, both without
\cite{Mazzitelli:2020jio} and with additional colour-singlet particles
\cite{Mazzitelli:2024ura}.

Within the \minnlo{} framework, NNLO QCD accuracy is achieved for observables that
are inclusive over QCD radiation, while retaining fully exclusive event generation.
The method thereby unifies higher-order accuracy with the parton-shower resummation
of logarithmically enhanced contributions. Its computational efficiency
and flexibility have enabled the construction of \minnlo{} generators for a wide
range of important LHC reactions \cite{Lombardi:2020wju,Lombardi:2021rvg,Mazzitelli:2020jio,
Mazzitelli:2021mmm,Buonocore:2021fnj,Lombardi:2021wug,Zanoli:2021iyp,
Gavardi:2022ixt,Haisch:2022nwz,Lindert:2022qdd,Mazzitelli:2023znt,
Mazzitelli:2024ura,Biello:2024vdh,Niggetiedt:2024nmp,Biello:2024pgo,
Biello:2026nhj}.

We briefly recall the \minnlo{} method for the production of a colour-singlet final state F (${\rm F}=HH$ in our case).
Starting from a \textsc{Powheg} \cite{Nason:2004rx,Frixione:2007vw,Alioli:2010xd} NLO+PS calculation for the production of a colour-singlet system in associated with an extra jet (FJ), the \minnlo{} master formula reads
\be
\text{d}\sigma_{\textrm{F}}^{\textsc{MiNNLO}_{\rm PS}} =
\text{d}\Phi_{\textrm{FJ}}\Bar{B}^{\textsc{MiNNLO}_{\rm PS}}
\times\left\{\Delta_{\textrm{pwg}}(\Lambda_{\textrm{pwg}})
+\text{d}\Phi_{\textrm{rad}}\Delta_{\textrm{pwg}}(p_{\textrm{T,rad}})
\frac{R_{\textrm{FJ}}}{B_{\textrm{FJ}}}\right\}\,,
\label{eq:minnlo}
\ee
where $B_{\textrm{FJ}}$ and $R_{\textrm{FJ}}$ are obtained from the squared tree-level matrix elements for FJ and FJJ production, respectively. The symbol $\Phi_{\textrm{FJ}}$ denotes the FJ phase space, $\Delta_{\textrm{pwg}}$ is the \textsc{Powheg} Sudakov form factor, with the default \textsc{Powheg} cutoff $\Lambda_{\rm pwg}=0.89$\,GeV, and $\Phi_{\textrm{rad}}$ and $p_{\textrm{T,rad}}$ represent the phase space and transverse momentum of the second emission.
The \textsc{Powheg} $\Bar{B}$ function is the central ingredient modified by the \minnlo{} method to achieve NNLO QCD accuracy for inclusive F production. It can be schematically written as
\be
\Bar{B}^{\textsc{MiNNLO}_{\rm PS}} = e^{-\tilde{S}}\left\{\frac{\text{d}\sigma_{\textrm{FJ}}^{(1)}}{\text{d}\Phi_{\textrm{FJ}}}\left(1+\tilde{S}^{(1)}\right)+\frac{\text{d}\sigma_{\textrm{FJ}}^{(2)}}{\text{d}\Phi_{\textrm{FJ}}}+\left(D-D^{(1)}-D^{(2)}\right)\times F^{\textrm{corr}}\right\}\,, \label{eq:btilde}
\ee
where $\text{d}\sigma_{\textrm{FJ}}^{(1,2)}$ denote the first- and second-order differential FJ cross sections, while $e^{-\tilde{S}}$ is the Sudakov form factor and $\tilde{S}(1)$ its $\mathcal{O}(\alpha_s)$ expansion.
Notice that the tilde indicates the modified resummation coefficients defined in the \minnlo{} formalism \cite{Monni:2019whf}.

This construction is based on the transverse-momentum ($\pt$) resummation formula
\be
\frac{\text{d}\sigma}{\text{d}\Phi_{\text{F}}\text{d}\pt} = \frac{\text{d}}{\text{d}\pt}\left\{e^{-\tilde{S}}\mathcal{L}\right\} = e^{-\tilde{S}}\left\{-\frac{\text{d}\tilde{S}}{\text{d}\pt}\mathcal{L}+\frac{\text{d}\mathcal{L}}{\text{d}\pt}\right\}\equiv e^{-\tilde{S}}D\,, \label{eq:pt_resum}
\ee
which defines the function $D$.
The luminosity factor $\mathcal{L}$ in \eqn{eq:pt_resum} includes the convolution of the parton distribution functions (PDF) with the collinear coefficient functions and the hard function, which comprises the 
squared virtual matrix elements of the process.

In the \minnlo{} procedure, the renormalisation and factorisation scales are set to
$\pt$, except for the overall powers of $\alpha_s$ appearing at Born level, whose
scale can be chosen independently. The last term in \eqn{eq:btilde}, starting at
relative order $\alpha_s^3(\pt)$, provides the singular contributions required to achieve
NNLO accuracy. Rather than truncating these terms at $\mathcal{O}(\alpha_s^3)$,
as in the original formulation~\cite{Monni:2019whf}, we follow the extension of
\cite{Monni:2020nks} and retain the full total-derivative structure in
\eqn{eq:pt_resum}. This preserves subleading logarithmic contributions beyond
$\mathcal{O}(\alpha_s^3)$ and improves the agreement with fixed-order NNLO
predictions. Finally, the factor $F^{\text{corr}}$ in \eqn{eq:btilde} distributes
the $D$ terms, which have Born-like kinematics, over the full FJ phase space.

\section{Approximations of top-quark mass effects}
\label{sec:topmass}

Radiative corrections to the $pp\to HH$ process are particularly challenging,
as the process is loop induced and involves multiple kinematic scales associated
with both internal and external masses. In particular, an NNLO calculation
requires three-loop corrections to the $2\to 2$ Born process and two-loop
corrections to the $2\to 3$ real-emission process. Their evaluation with full mass dependence is currently beyond the reach of existing techniques.

As a consequence, the amplitudes required for a complete NNLO calculation in the
full theory, including exact top-quark mass effects, are not known. At present,
only the one-loop amplitudes and the two-loop $gg\to HH$ contribution are
available with full mass dependence. In order to  construct approximate NNLO predictions, we
therefore exploit the available information to define a set of approximations
that incorporate top-quark mass effects beyond NLO accuracy.
In the following, we introduce these approximations, whose phenomenological
impact will be investigated in \sct{sec:results_mass}.

The first and most basic approximation we consider is the heavy-top limit, 
in which the top quark is assumed to be infinitely heavy and integrated out. 
In this framework, each top-quark loop in \fig{fig:feyn_diagrams} is replaced 
by an effective Higgs--gluon interaction. 
The tree-level amplitudes for $pp\to HHj$ and $pp\to HHjj$ in the HTL, 
as well as the one-loop amplitude for $pp\to HHj$, are computed using 
\ol~\cite{Cascioli:2011va,Buccioni:2017yxi,Buccioni:2019sur}. 
The $gg\to HH$ amplitudes up to two loops are instead implemented analytically 
using the results of \citere{deFlorian:2013jea,deFlorian:2016uhr}. The 
relevant expressions are collected in \app{app:H1_H2_ggHH}.\footnote{We have validated 
the analytic tree-level and one-loop results against \ol.}

We construct improved approximations by progressively incorporating information available in the full theory (FT).
The simplest approach to include top-mass effects consists in rescaling the differential LO prediction in the full theory by higher-order corrections computed in the HTL. This defines the LO-improved (\loi{}) approximation
\begin{align}
{\rm d} \sigma_{\rm\loi}^{\rm NNLO} =
{\rm d} \sigma_{\rm FT}^{\rm LO}
\cdot
\frac{{\rm d} \sigma_{\rm HTL}^{\rm NNLO}}{{\rm d} \sigma_{\rm HTL}^{\rm LO}}\,,
\end{align}
which is obtained by computing all contributions separately and combining them \emph{a posteriori} at histogram level.
Since the NLO cross section is also known in the full theory, this procedure can be extended to include exact top-mass effects at NLO, leading to the NLO-improved (\nloi{}) approximation
\begin{align}
{\rm d} \sigma_{\rm\nloi}^{\rm NNLO} =
{\rm d} \sigma_{\rm FT}^{\rm NLO}
\cdot
\frac{{\rm d} \sigma_{\rm HTL}^{\rm NNLO}}{{\rm d} \sigma_{\rm HTL}^{\rm NLO}}\,.
\end{align}

While conceptually simple, such approaches are not well suited for Monte Carlo event generation. They require the production of multiple independent event samples that can only be combined at the level of observables, thereby preventing a fully exclusive and self-contained description. In contrast, a {\it local} implementation of mass effects at the level of matrix elements, i.e. {\it point-by-point} in phase space, is preferable. This allows one to retain a single, fully exclusive event sample and generally provides a more physical description of the underlying dynamics.

For this reason, we introduce a set of approximations defined directly at the level of the finite remainders $|\mathcal R\rangle$ of the relevant amplitudes.\footnote{The finite remainder is defined following \citeres{Becher:2009qa,Becher:2010tm} throughout this paper.}
We start from the simplest possibility, namely a global rescaling of all HTL amplitudes by the ratio of the full-theory and HTL Born squared matrix elements
\begin{align}
R =
\frac{\langle\mathcal{R}^{(0)}_{HH}|\mathcal{R}^{(0)}_{HH}\rangle_{\rm FT}}
{\langle\mathcal{R}^{(0)}_{HH}|\mathcal{R}^{(0)}_{HH}\rangle_{\rm HTL}}\,,
\end{align}
which defines the Born-improved (\borni{}) approximation:
\begin{align}
\begin{split}\label{born-I}
\langle \mathcal{R}_{HH}^{{(0)}} |\mathcal{R}_{HH}^{{(0)}}\rangle_{\rm \borni} &= \langle \mathcal{R}_{HH}^{{(0)}} |\mathcal{R}_{HH}^{{(0)}}\rangle_{\rm FT}\,,\\
2\,{\rm Re}\langle \mathcal{R}_{HH}^{{(0)}} |\mathcal{R}_{HH}^{{(1)}}\rangle_{\rm \borni} &= 2\,{\rm Re}\langle\mathcal{R}^{(0)}_{HH} |\mathcal{R}^{(1)}_{HH}\rangle_{\rm HTL} \cdot R\,,\\
\langle \mathcal{R}_{HH}^{{(1)}} |\mathcal{R}_{HH}^{{(1)}}\rangle_{\rm \borni} &= \langle\mathcal{R}^{(1)}_{HH} |\mathcal{R}^{(1)}_{HH}\rangle_{\rm HTL} \cdot R\,,\\
2\,{\rm Re}\langle \mathcal{R}_{HH}^{{(0)}} |\mathcal{R}_{HH}^{{(2)}}\rangle_{\rm \borni} &= 2\,{\rm Re}\langle\mathcal{R}^{(0)}_{HH} |\mathcal{R}^{(2)}_{HH}\rangle_{\rm HTL} \cdot R\,,\\[0.2cm]
\langle \mathcal{R}_{HH{\rm J}}^{{(0)}} |\mathcal{R}_{HH{\rm J}}^{{(0)}}\rangle_{\rm \borni} &= \langle\mathcal{R}^{(0)}_{HH{\rm J}} |\mathcal{R}^{(0)}_{HH{\rm J}}\rangle_{\rm HTL} \cdot R\,,\\
2\,{\rm Re}\langle \mathcal{R}_{HH{\rm J}}^{{(0)}} |\mathcal{R}_{HH{\rm J}}^{{(1)}}\rangle_{\rm \borni} &= 2\,{\rm Re}\langle\mathcal{R}^{(0)}_{HH{\rm J}} |\mathcal{R}^{(1)}_{HH{\rm J}}\rangle_{\rm HTL} \cdot R\,,\\[0.2cm]
\langle \mathcal{R}_{HH{\rm JJ}}^{{(0)}} |\mathcal{R}_{HH{\rm JJ}}^{{(0)}}\rangle_{\rm \borni} &= \langle\mathcal{R}^{(0)}_{HH{\rm JJ}} |\mathcal{R}^{(0)}_{HH{\rm JJ}}\rangle_{\rm HTL} \cdot R\,.
\end{split}
\end{align}
For amplitudes with additional radiation ($HH{\rm J}$ and $HH{\rm JJ}$), the ratio $R$ is evaluated using the standard \POWHEG{} projection of the 
corresponding kinematics onto the $HH$ phase space.

We now introduce three refined approximations, in which progressively more information from the full theory is incorporated. The one denoted as FT-approx has been previously introduced in the context of a NNLO fixed-order calculation \cite{Grazzini:2018bsd}, and we refer to the other two as FT-approx-0 and FT-approx-2. In all cases, we use the exact full-theory Born and real-emission contributions, while approximating only the virtual corrections.
In particular, the virtual corrections to $HH{\rm J}$ production are treated in all three approximations as
\begin{align}
2\,{\rm Re}\langle \mathcal{R}_{HH{\rm J}}^{{(0)}} |\mathcal{R}_{HH{\rm J}}^{{(1)}}\rangle_{\textrm{FT-approx}} &= 2\,{\rm Re}\langle\mathcal{R}^{(0)}_{HH{\rm J}} |\mathcal{R}^{(1)}_{HH{\rm J}}\rangle_{\rm HTL} \cdot \frac{\langle\mathcal{R}^{(0)}_{HH{\rm J}} |\mathcal{R}^{(0)}_{HH{\rm J}}\rangle_{\rm FT}}{\langle\mathcal{R}^{(0)}_{HH{\rm J}} |\mathcal{R}^{(0)}_{HH{\rm J}}\rangle_{\rm HTL}}\,,
\end{align}
while they differ in their treatment of the virtual $gg\to HH$ amplitudes:
\begin{itemize}
\item In FT-approx-0, both the first- and second-order virtual contributions are obtained by a Born-level rescaling of the HTL results,
\begin{align}
\begin{split}
2\,{\rm Re}\langle \mathcal{R}_{HH}^{{(0)}} |\mathcal{R}_{HH}^{{(1)}}\rangle_{\textrm{FT-approx-}0} &= 2\,{\rm Re}\langle\mathcal{R}^{(0)}_{HH} |\mathcal{R}^{(1)}_{HH}\rangle_{\rm HTL} \cdot R\,,\\
\langle \mathcal{R}_{HH}^{{(1)}} |\mathcal{R}_{HH}^{{(1)}}\rangle_{\textrm{FT-approx-}0} &= \langle\mathcal{R}^{(1)}_{HH} |\mathcal{R}^{(1)}_{HH}\rangle_{\rm HTL} \cdot R\,,\\
2\,{\rm Re}\langle \mathcal{R}_{HH}^{{(0)}} |\mathcal{R}_{HH}^{{(2)}}\rangle_{\textrm{FT-approx-}0} &= 2\,{\rm Re}\langle\mathcal{R}^{(0)}_{HH} |\mathcal{R}^{(2)}_{HH}\rangle_{\rm HTL} \cdot R\,.
\label{eq:aFT1}
\end{split}
\end{align}

\item In FT-approx, the exact full-theory result is used for the two-loop virtual amplitude, while the three-loop contribution is still approximated as in \textrm{FT-approx-}0,
\begin{align}
  \begin{split}
 2\,{\rm Re}\langle \mathcal{R}_{HH}^{{(0)}} |\mathcal{R}_{HH}^{{(1)}}\rangle_{\rm  \textrm{FT-approx}} &= 2\,{\rm Re}\langle\mathcal{R}^{(0)}_{HH} |\mathcal{R}^{(1)}_{HH}\rangle_{\rm FT} \,, \\    
\langle \mathcal{R}_{HH}^{{(1)}} |\mathcal{R}_{HH}^{{(1)}}\rangle_{\rm\textrm{FT-approx}} &= \langle\mathcal{R}^{(1)}_{HH} |\mathcal{R}^{(1)}_{HH}\rangle_{\rm HTL} \cdot R\,,\\
2\,{\rm Re}\langle \mathcal{R}_{HH}^{{(0)}} |\mathcal{R}_{HH}^{{(2)}}\rangle_{\rm  \textrm{FT-approx}} &= 2\,{\rm Re}\langle\mathcal{R}^{(0)}_{HH} |\mathcal{R}^{(2)}_{HH}\rangle_{\rm HTL} \cdot R\,.
\end{split}
\end{align}

\item Finally, in FT-approx-2, the available two-loop information in the full theory is used to improve the approximation of the three-loop contribution,
\begin{align}
  \begin{split}
2\,{\rm Re}\langle \mathcal{R}_{HH}^{{(0)}} |\mathcal{R}_{HH}^{{(1)}}\rangle_{\textrm{FT-approx-}2} & = 2\,{\rm Re}\langle\mathcal{R}^{(0)}_{HH} |\mathcal{R}^{(1)}_{HH}\rangle_{\rm FT} \,, \\        
\langle \mathcal{R}_{HH}^{{(1)}} |\mathcal{R}_{HH}^{{(1)}}\rangle_{\textrm{FT-approx-}2} &= \langle\mathcal{R}^{(1)}_{HH} |\mathcal{R}^{(1)}_{HH}\rangle_{\rm HTL}  \cdot \frac{2\,{\rm Re}\langle\mathcal{R}^{(0)}_{HH} |\mathcal{R}^{(1)}_{HH}\rangle_{\rm FT}}{2\,{\rm Re}\langle\mathcal{R}^{(0)}_{HH} |\mathcal{R}^{(1)}_{HH}\rangle_{\rm HTL}}\,,\\
2\,{\rm Re}\langle \mathcal{R}_{HH}^{{(0)}} |\mathcal{R}_{HH}^{{(2)}}\rangle_{\textrm{FT-approx-}2} &= 2\,{\rm Re}\langle\mathcal{R}^{(0)}_{HH} |\mathcal{R}^{(2)}_{HH}\rangle_{\rm HTL} \cdot \frac{2\,{\rm Re}\langle\mathcal{R}^{(0)}_{HH} |\mathcal{R}^{(1)}_{HH}\rangle_{\rm FT}}{2\,{\rm Re}\langle\mathcal{R}^{(0)}_{HH} |\mathcal{R}^{(1)}_{HH}\rangle_{\rm HTL}}\,.
\end{split}
\end{align}
\end{itemize}

For the full-theory amplitudes, all one-loop contributions are evaluated using \ol{}.
For the two-loop $gg\to HH$ amplitude, we employ two independent implementations. The original numerical calculation is available in the form of interpolation grids provided by \textsc{HHgrid} \cite{Borowka:2016ehy,Borowka:2016ypz,Heinrich:2017kxx,Davies:2019dfy}. However, due to the limited density of grid points, sizeable fluctuations can arise in sparsely populated regions of phase space.
To improve numerical stability, we also use the \textsc{ggxy} library~\cite{Davies:2025qjr}, which is based on semi-analytic expansions. In particular, at small transverse momenta of the Higgs pair we observe significant fluctuations when using \textsc{HHgrid}, likely due to interpolation artefacts.
For this reason, we adopt \textsc{ggxy} as the default choice for our predictions. We have verified that both implementations agree within statistical uncertainties in regions where \textsc{HHgrid} is sufficiently populated.

\FloatBarrier

\section{Results with stable Higgs bosons}
\label{sec:results}
In this section we present phenomenological results for Higgs-boson pair production at the LHC. We focus on on-shell Higgs bosons, while the inclusion of Higgs decays is deferred to \sct{sec:higgsdecays}. We adopt the following input parameters:
\be
m_H = 125\,\GeV, \quad G_F = 1.663787\times 10^{-5}\,\GeV^{-2}, \quad m_t = 173\,\GeV, \quad \Gamma_H = \Gamma_t = 0\,.
\label{eq:input_par}
\ee
We use $n_f=5$ massless quark flavours and the PDF4LHC21\_40 PDF set
(LHAPDF ID 93100). The renormalisation scale associated with the two
overall powers of $\alpha_s$ is set to $K_R\,m_{HH}$, where $m_{HH}$
denotes the invariant mass of the Higgs pair. The renormalisation scale
for additional powers of $\alpha_s$ and the factorisation scale are
chosen according to the standard \minnlo{} setup. Unless stated
otherwise, results are presented for a centre-of-mass energy of
$13\,\TeV$, with central scale choices $K_R=K_F=0.5$. Perturbative
uncertainties are estimated through seven-point scale variations,
varying $K_R$ and $K_F$ independently by a factor of two up and down,
with the constraint $1/2\leq K_R/K_F\leq 2$. All results, except those
in \sct{sec:matrix}, have been showered using
\textsc{Pythia}8~\cite{Sjostrand:2014zea}. Since we don't compare our results with data, we turned off the QED shower, hadronisation and multi-parton interaction (MPI) effects, but they can be easily included if needed.

\subsection{Comparison with fixed-order NNLO QCD predictions}
\label{sec:matrix}
We start by validating our results against fixed-order NNLO QCD predictions
obtained with \textsc{Matrix}~\cite{Grazzini:2017mhc,Grazzini:2018bsd}, comparing both results in FT-approx.
The setup is the same as described above, except that we choose a
centre-of-mass energy of $14\,\TeV$ and use the PDF4LHC15\_nnlo\_100 PDF set
(LHAPDF ID 91700), in order to match the input parameters adopted for the
\textsc{Matrix} results of \citere{Grazzini:2018bsd}. For a more direct
comparison with the fixed-order calculation, we use results at lhe level.

\begin{figure}[t]
    \centering
    \includegraphics[width=0.47\textwidth]{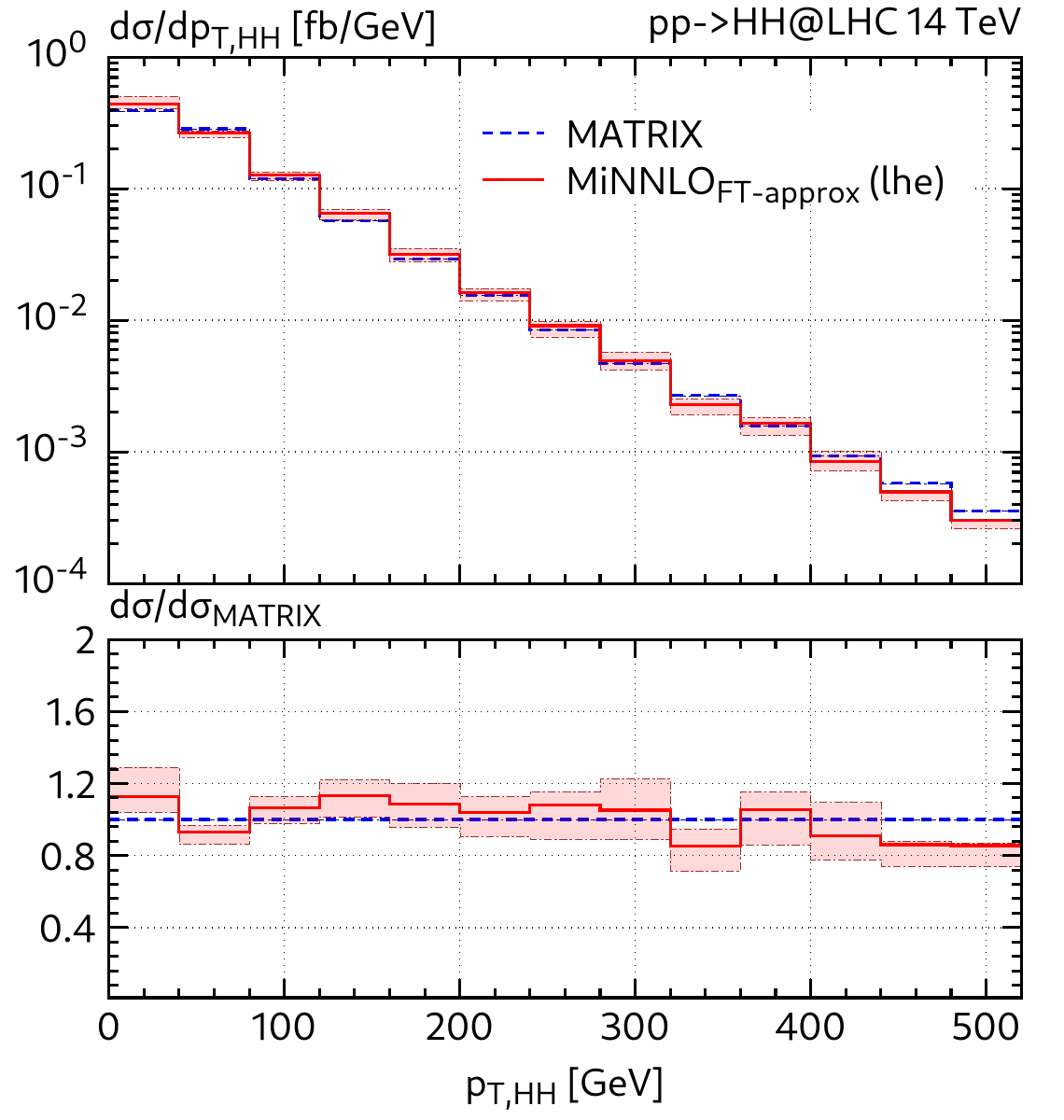}
    \includegraphics[width=0.47\textwidth]{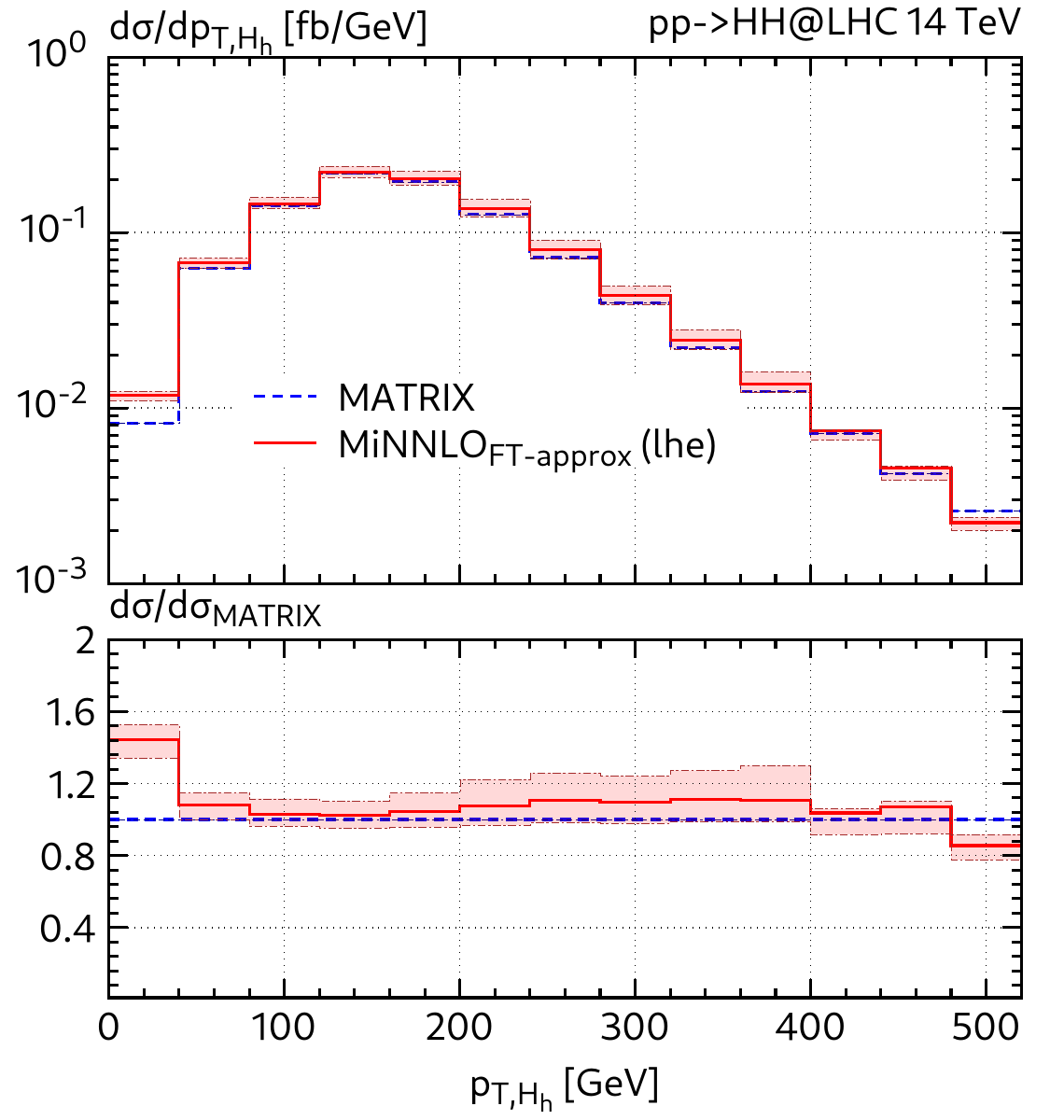}
    \includegraphics[width=0.47\textwidth]{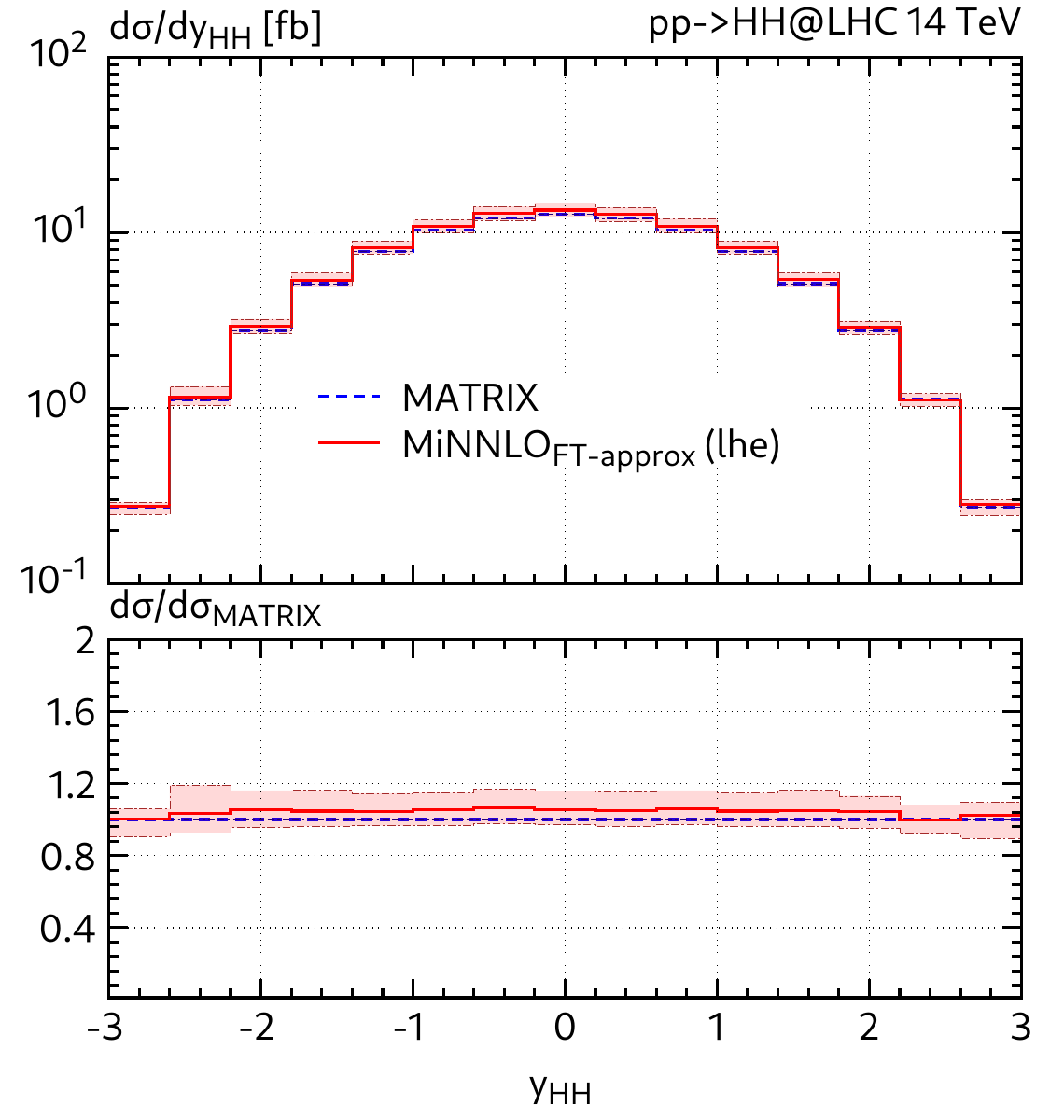}
    \includegraphics[width=0.47\textwidth]{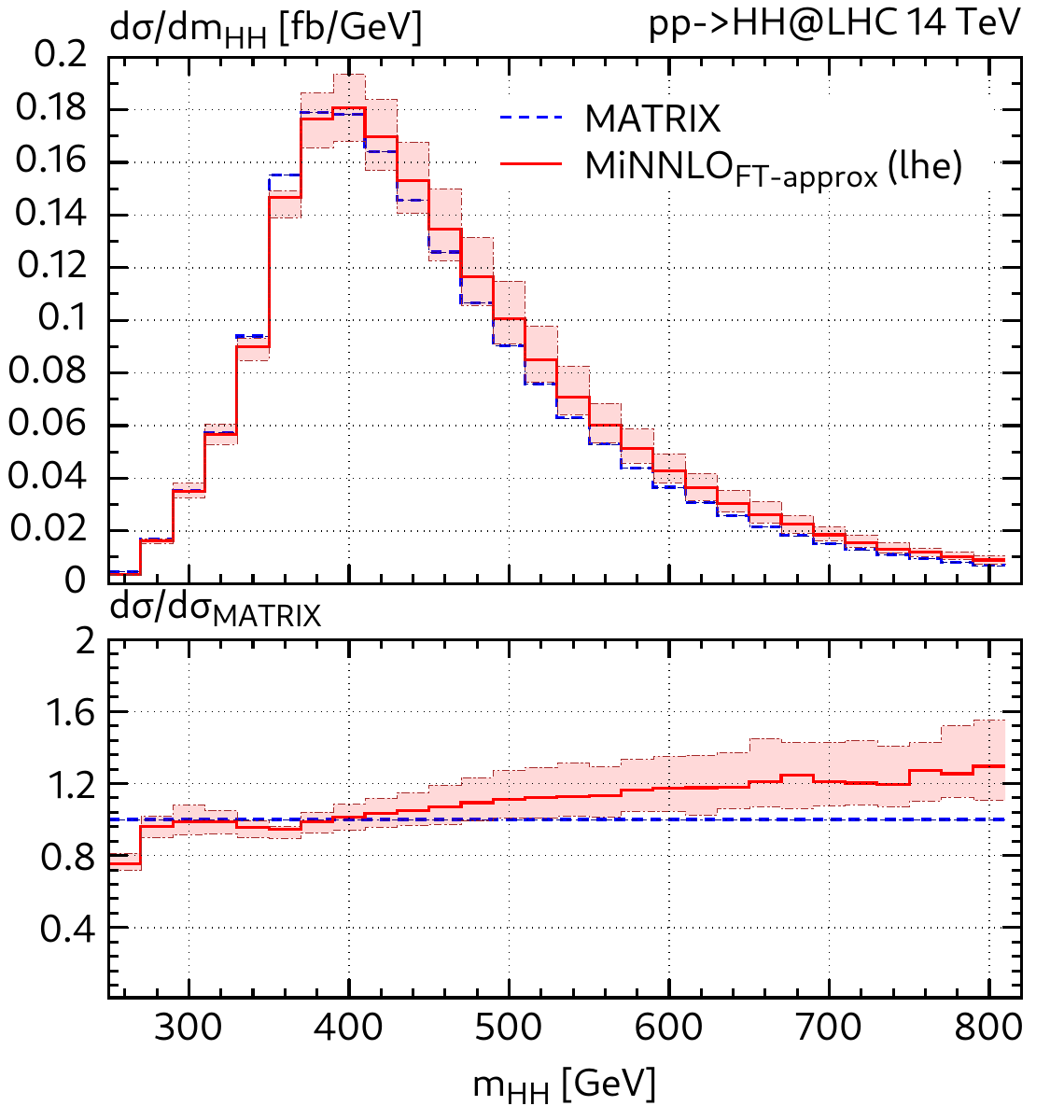}
    \caption{Comparison of \minnlo{} predictions (red)
    against the fixed-order results (blue), both in FT-approx. We show distributions for the transverse
    momentum of the Higgs pair (top-left), the transverse momentum of the hardest
    Higgs boson (top-right), the rapidity of the Higgs pair (bottom-left), and its
    invariant mass (bottom-right).}
    \label{fig:matrix}
\end{figure}

In \fig{fig:matrix} we present the comparison for four differential
observables: the transverse momentum of the Higgs pair
($p_{\textrm{T},HH}$), the transverse momentum of the hardest Higgs boson
($p_{\textrm{T},H_h}$), the rapidity ($y_{HH}$), and the invariant mass
($m_{HH}$) of the Higgs pair. We note that $p_{\textrm{T},HH}$ requires
recoil against QCD radiation and is therefore effectively only NLO
accurate, being particularly sensitive to soft and collinear emissions.
Despite this, we observe good agreement across all observables. This
provides a non-trivial validation of the two implementations, especially
regarding the treatment of top-quark mass effects, which—as will be
discussed in the comparison with \textsc{Geneva} in
\sct{sec:geneva}—can lead to sizeable differences.

Only the large-$m_{HH}$ region exhibits somewhat more pronounced differences.
A quantitative comparison is limited by the fact that only central
predictions were available to us from \citere{Grazzini:2018bsd},
precluding a more meaningful comparison including uncertainties. In
particular, differences at large $m_{HH}$ can arise from configurations
with relatively small $p_{\textrm{T},HH}$, where resummation effects may
become relevant and fixed-order predictions are less reliable.

\subsection{Mass effects}
\label{sec:results_mass}
We now turn to the impact of finite top-quark mass effects in
Higgs-boson pair production. As a first step, we compare in
\tab{tab:sigma_13TeV_nlo} different approximations for the total cross
section at NLO against the full-theory prediction. The \borni{}
result corresponds to the approximation in \eqn{born-I} with the NNLO
contributions omitted, while FT-approx denotes the 
approximation of \eqn{eq:aFT1}, again restricted to NLO accuracy. We
observe that finite top-mass effects increase the NLO cross section by
about $7\%$, similarly to the case of single-Higgs production. The
\borni{} approximation substantially overestimates these effects,
yielding a cross section roughly $20\%$ larger than the exact NLO
result, whereas the FT-approx result reproduces the full-theory
prediction with very good accuracy. This provides a robust validation of
the FT approximation at NLO and supports the use of analogous
approximations also at NNLO.

We continue by presenting NNLO predictions for the total inclusive cross section
in the various approximations introduced in \sct{sec:topmass}. The
corresponding results are collected in \tab{tab:sigma_13TeV}. We
observe that the HTL prediction yields the smallest cross section,
reflecting sizeable positive finite-mass effects. One of the best-motivated approaches to
incorporate mass effects in inclusive NNLO observables is the \nloi{}
approximation, which assumes that the NNLO/NLO $K$-factor obtained in
the HTL can be applied to the full-theory NLO prediction. However, since this procedure is defined
only at the level of differential cross sections and applied
{\it a posteriori}, it is less suitable in the context of fully
exclusive event generation, as discussed in detail in  \sct{sec:topmass}.

By contrast, the three full-theory approximations (FT-approx-0, FT-approx and
FT-approx-2) implement mass
effects locally in phase space at the matrix-element level. We
find that all three approximations are in excellent agreement with the
\nloi{} prediction and with each other. Their total cross sections differ by 
less than $1.5\%$, well below the corresponding scale
uncertainties. This indicates not only that these approaches provide reliable
approximations of finite top-mass effects, but also that the residual
ambiguity associated with the treatment of the virtual $HH$ amplitudes
induces only a very moderate uncertainty. The \borni{}
prediction, however, is substantially larger, suggesting that a simple
Born-level rescaling tends to overestimate the impact of finite
top-mass effects on the inclusive cross section. While even the HTL
prediction remains compatible with the FT approximations within scale
uncertainties, this is no longer the case for the \borni{} prediction.

\begin{table}[t]
\begin{center}
\begin{tabular}{| c | c | c | c | }
\hline
 HTL & \borni{} & FT-approx & NLO \\
\hline
 \rule[-1.5ex]{0pt}{5ex} $25.73^{+17,7\%}_{-15.3\%}$ & $32.81^{+18.1\%}_{-14.9\%}$ & $28.44^{+14.8\%}_{-13.3\%}$ & $27.35^{+13.7\%}_{-12.7\%}$ \\
\hline
\end{tabular}
\caption{Total NLO cross sections (in fb) at $13\,\TeV$ in three approximations compared to the exact NLO result.\label{tab:sigma_13TeV_nlo}}
\end{center} 
\end{table}

\begin{table}[t]
\begin{center}
\begin{tabular}{ | c | c | c | c | c | c | }
\hline
HTL & \borni{} & \nloi{} & \textrm{FT-approx-}0 & \textrm{FT-approx} & \textrm{FT-approx-}2 \\
\hline
\rule[-1.5ex]{0pt}{5ex} $30.08^{+8.1\%}_{-8.5\%}$ & $37.34^{+9.2\%}_{-8.7\%}$ & $31.98^{+4.4\%}_{-5.8\%}$ & $32.04^{+6.1\%}_{-6.6\%}$ & $32.32^{+9.5\%}_{-8.1\%}$ & $31.88^{+8.6\%}_{-7.7\%}$ \\
\hline
\end{tabular}
\caption{Total \minnlo{} cross sections (in fb) in the six approximations at $13\,\TeV$.\label{tab:sigma_13TeV}}
\end{center} 
\end{table}

\begin{figure}[t]
    \centering
    \includegraphics[width=0.47\textwidth]{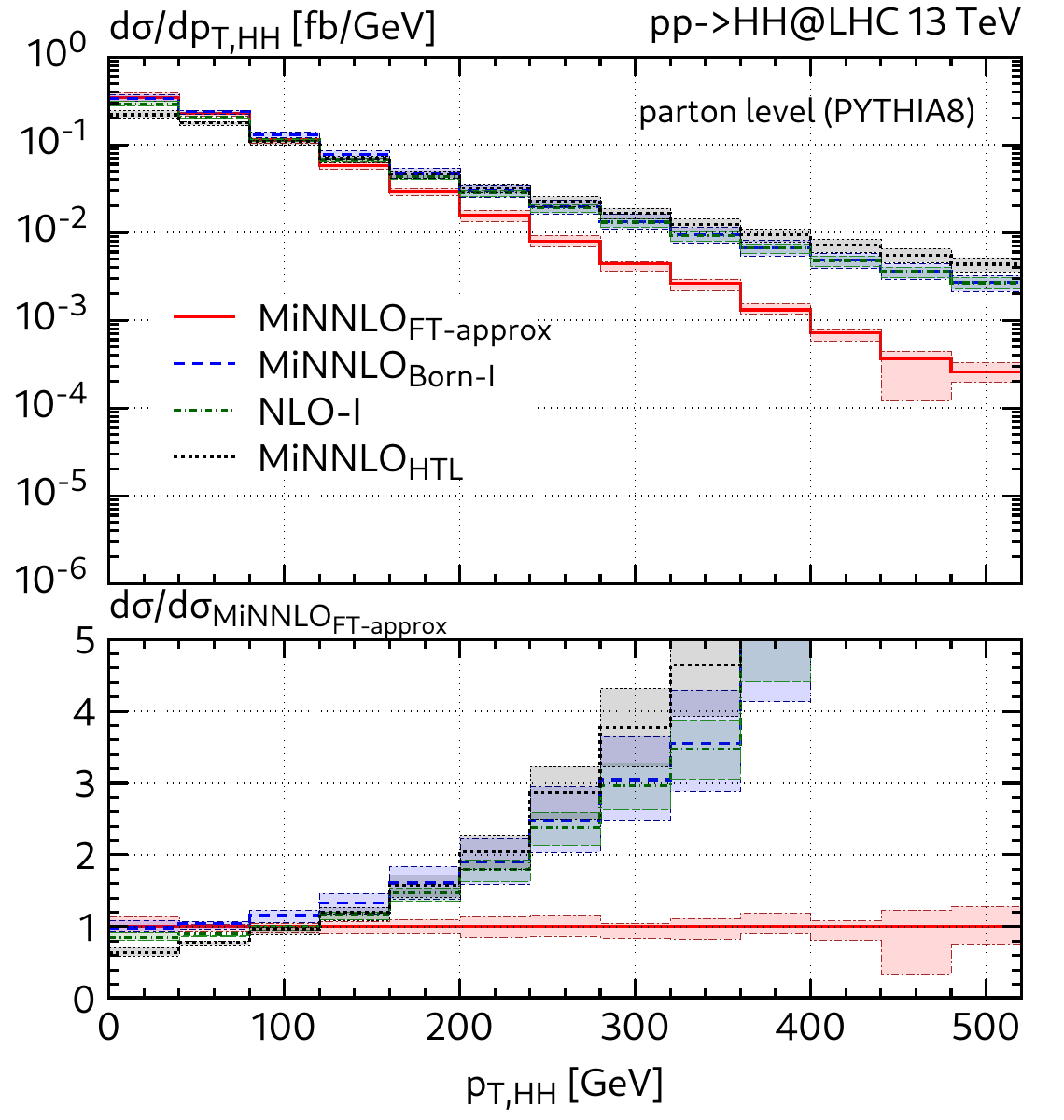}
    \includegraphics[width=0.47\textwidth]{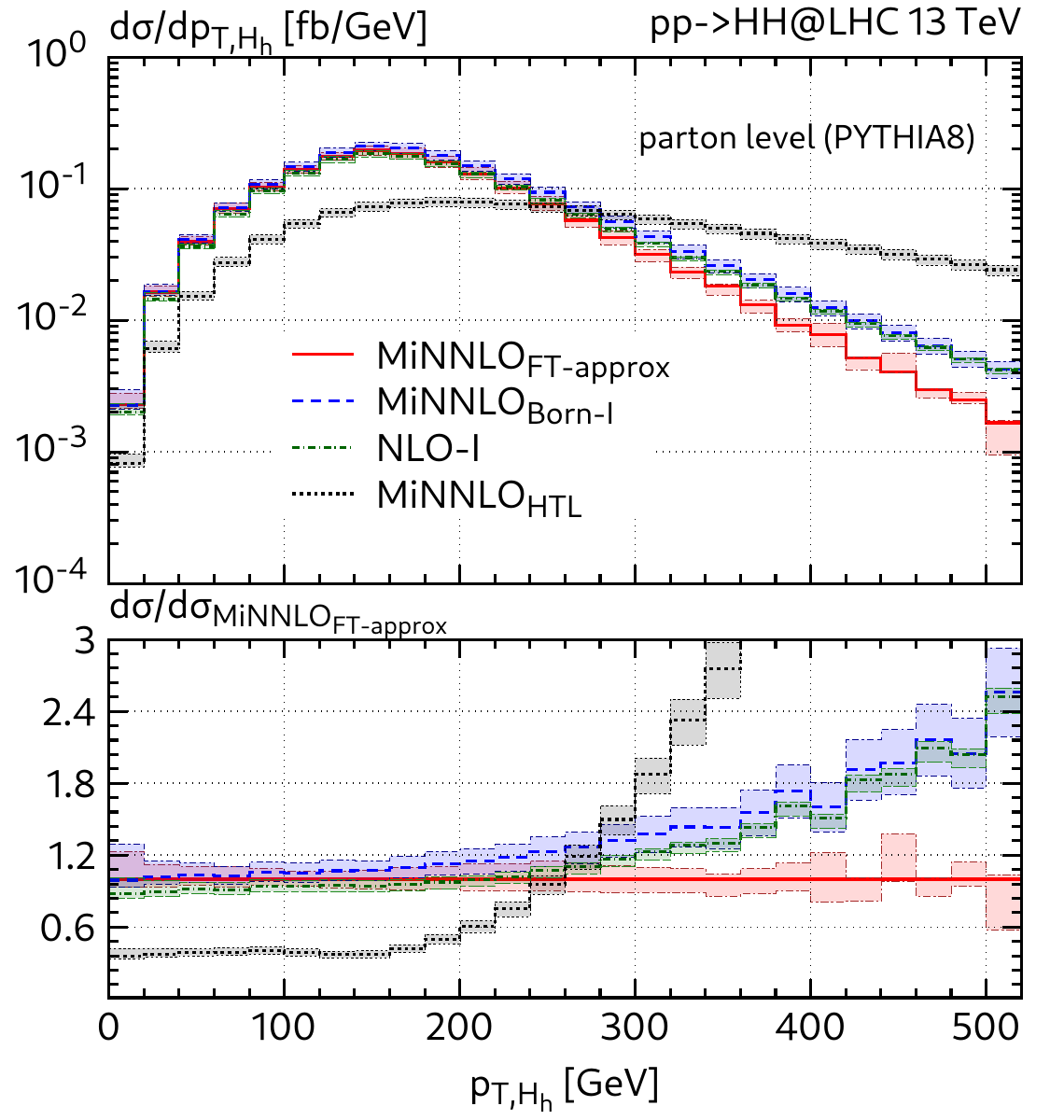}
    \includegraphics[width=0.47\textwidth]{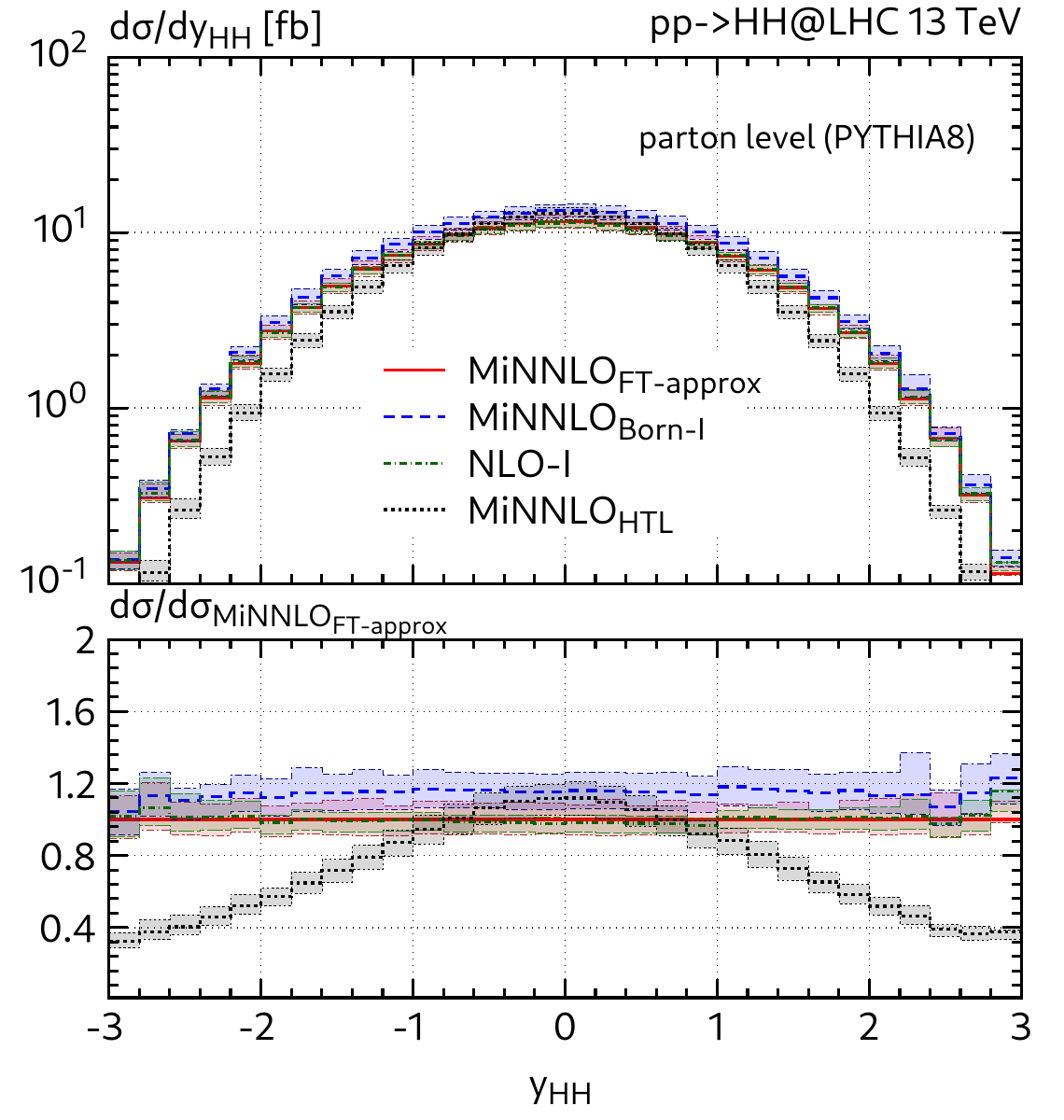}
    \includegraphics[width=0.47\textwidth]{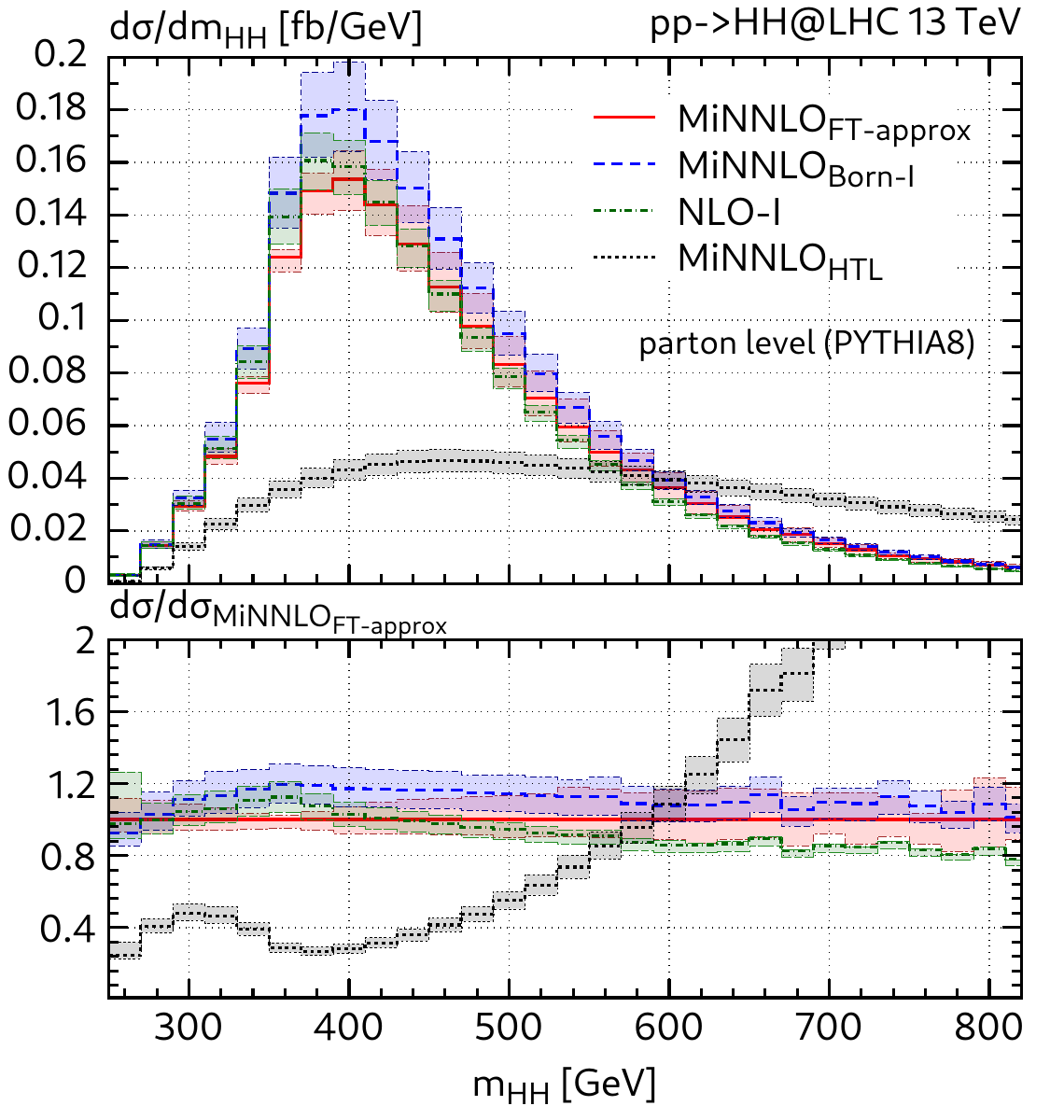}
    \caption{Comparison of \minnlo{} predictions in the FT-approx (red), \borni{} (blue), \nloi{} (green) and HTL (black)
    approximations. We show distributions for the transverse momentum
    of the Higgs pair (top-left), the transverse momentum of the
    hardest Higgs boson (top-right), the rapidity of the Higgs pair
    (bottom-left), and the invariant mass of the Higgs pair
    (bottom-right).}
    \label{fig:htl_bi_ft}
\end{figure}

In \fig{fig:htl_bi_ft} we compare predictions obtained in the FT-approx, HTL,
\borni{} and \nloi{} approximations for four representative
observables: the transverse momentum of the Higgs pair, the transverse
momentum of the hardest Higgs boson, the rapidity and the invariant mass
of the Higgs pair. We observe that the HTL approximation fails to
provide a reliable physical description, in particular regarding the
shapes of the distributions. The disagreement is less pronounced in the
central rapidity region, which also explains the comparatively reasonable
description of the inclusive cross section, but becomes substantially
larger at forward rapidities.

Finite top-quark mass effects are very sizeable, in particular at the
differential level. It is interesting to observe that all approximations
including mass effects, namely \borni{}, \nloi{} and FT-approx, lead to
rather similar predictions for observables inclusive over QCD radiation,
such as $m_{HH}$ and $y_{HH}$. By contrast, substantial differences
emerge at large transverse momenta, in particular in the
$p_{\textrm{T},HH}$ and $p_{\textrm{T},H_h}$ distributions, which are
directly sensitive to hard QCD recoil. In this region, only the
FT-approx provides a reliable physical description, since it
includes the exact double-real matrix elements in the full theory.

\begin{figure}[t]
    \centering
    \includegraphics[width=0.47\textwidth]{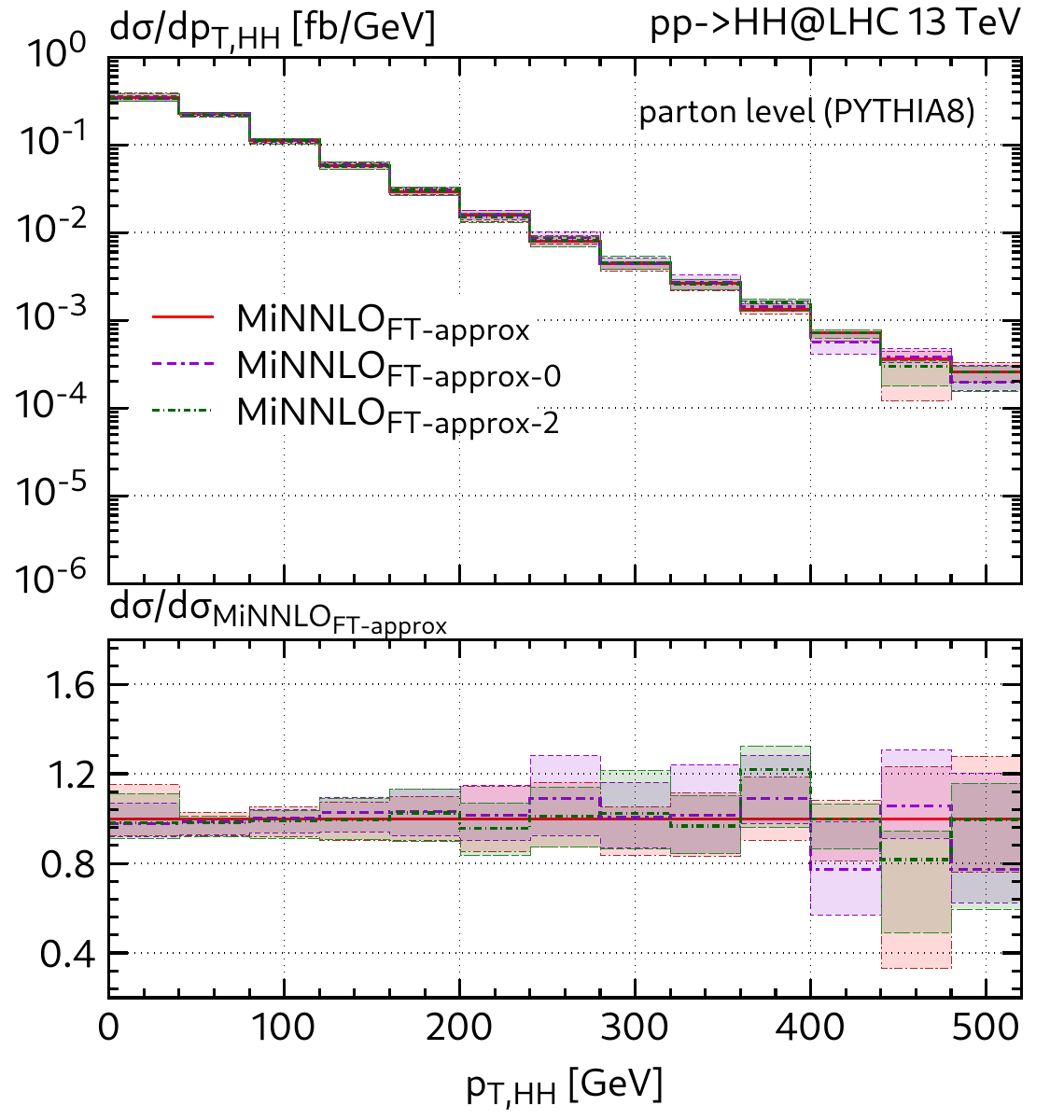}
    \includegraphics[width=0.47\textwidth]{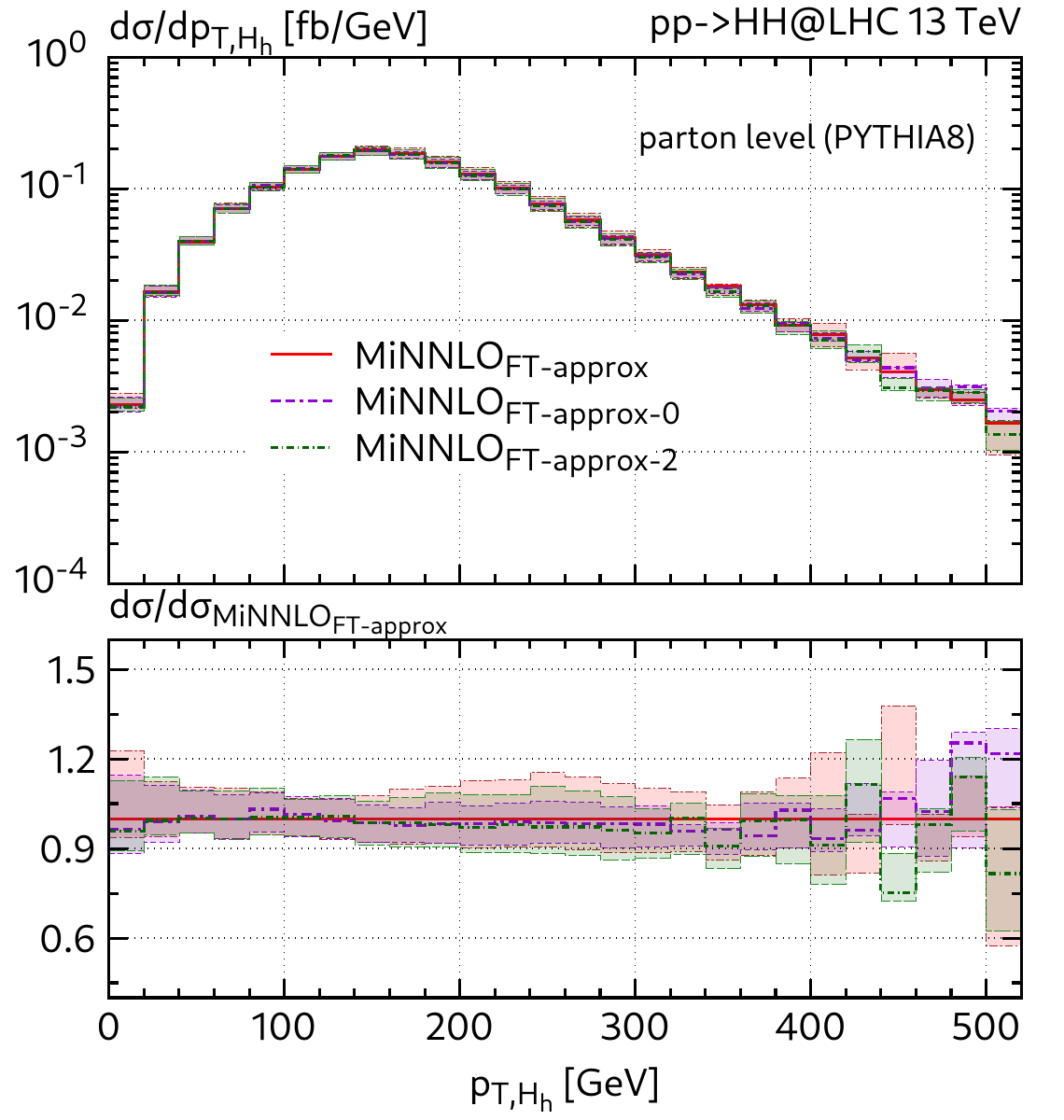}
    \includegraphics[width=0.47\textwidth]{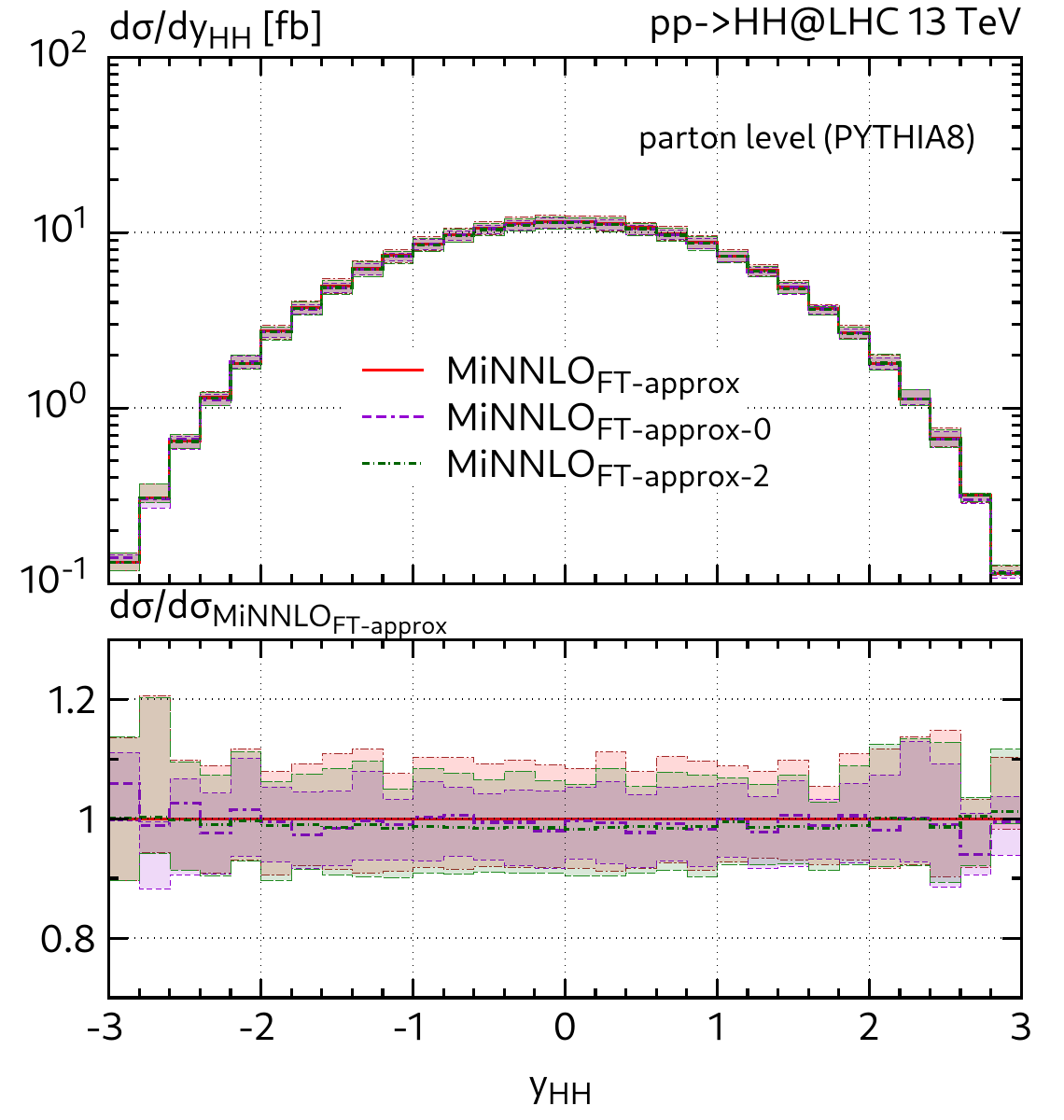}
    \includegraphics[width=0.47\textwidth]{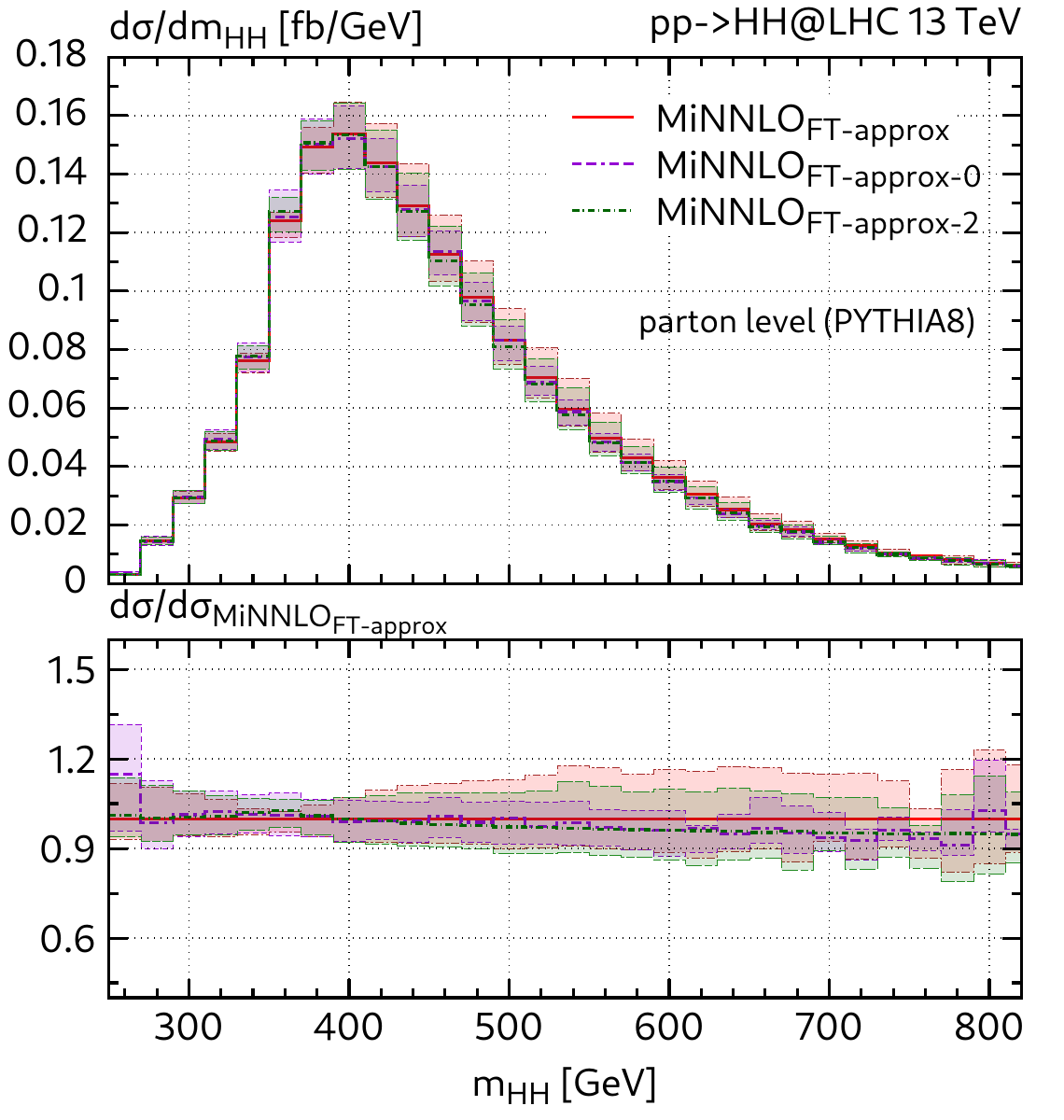}
    \caption{Comparison of the \minnlo{} predictions in the three FT
    approximations. We show distributions for the transverse momentum
    of the Higgs pair (top-left), the transverse momentum of the
    hardest Higgs boson (top-right), the rapidity of the Higgs pair
    (bottom-left), and the invariant mass of the Higgs pair
    (bottom-right).}
    \label{fig:ft123}
\end{figure}

The limitations of the \borni{} and \nloi{} approximations can be traced
back to their treatment of additional QCD radiation. In the
\borni{} approximation, all matrix elements are rescaled solely through
the underlying $HH$ kinematics, such that jet-sensitive observables
remain effectively described by the HTL. In the \nloi{} case, the
high-$p_{\textrm{T},HH}$ region, and consequently also the tail of
$p_{\textrm{T},H_h}$, is described with full top-mass dependence only at
leading order, while higher-order corrections are inherited from the
HTL. As a consequence, only the FT-approx yields a
consistent description of observables sensitive to hard QCD radiation.
The differences at large $p_{\textrm{T},H_h}$ arise from configurations
in which one Higgs boson recoils against a hard jet, while the second
Higgs remains comparatively soft. 
Our findings concerning the impact of top-quark mass effects in the different approximations are consistent with those of \citere{Alioli:2025xcu}, sections 4 and 5.

In \fig{fig:ft123}, we compare the predictions in the three FT approximations for
the same set of observables. We observe only very small differences
among the three predictions, with all results remaining compatible
within their respective uncertainty bands across the full kinematic
range considered. This indicates that the residual uncertainty
associated with the approximate treatment of top-quark mass effects
beyond the exactly known contributions remains moderate.

\subsection{Comparison with \textsc{Geneva}}
\label{sec:geneva}

\begin{table}[b]
\begin{center}
\begin{tabular}{ | c | c | c | c | }
\hline
$\sigma$ (fb) & HTL & \borni{} & FT-approx \\
\hline
\rule[-1.5ex]{0pt}{5ex} \minnlo{} & $31.32^{+9.2\%}_{-9.4\%}$ & $37.92^{+9.3\%}_{-9.2\%}$ & $32.72^{+7.7\%}_{-8.5\%}$ \\
\hline
\rule[-1.5ex]{0pt}{5ex} \textsc{Geneva} & $30.69^{+7.1\%}_{-8.5\%}$ & $37.65^{+7.3\%}_{-8.4\%}$ & $32.34^{+6.8\%}_{-5.3\%}$ \\
\hline
\end{tabular}
\caption{Total cross sections (in fb) in the HTL, \borni{} and FT-approx approximations at $13.6\,\TeV$ centre-of-mass energy for \minnlo{} and \textsc{Geneva} \cite{Alioli:2025xcu}. \label{tab:sigma_geneva}}
\end{center} 
\end{table}

We continue by comparing our results with the corresponding predictions
from \textsc{Geneva}~\cite{Alioli:2025xcu} at parton-shower level. To match
their setup, we work at a centre-of-mass energy of $13.6\,\TeV$ and use
central scale choices $K_R=K_F=1$. The corresponding total cross
sections in the HTL, \borni{} and FT-approx setups are collected in
\tab{tab:sigma_geneva}. We find good agreement within the respective
scale uncertainties in all cases. Since the two NNLO+PS frameworks treat
contributions beyond their formal accuracy differently, exact agreement
is not expected.

\begin{figure}[p!]
    \centering
    \includegraphics[width=0.47\textwidth]{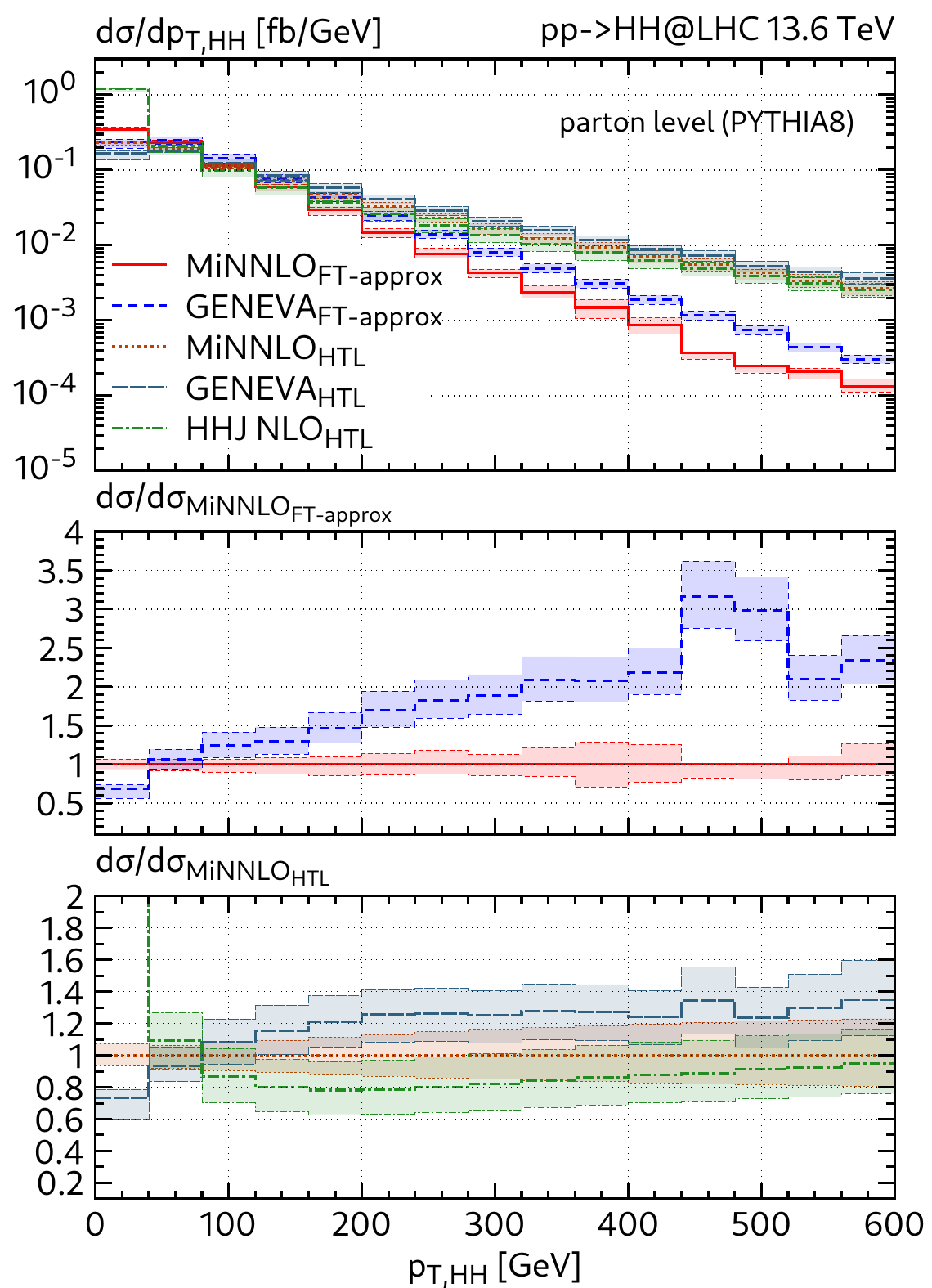}
    \includegraphics[width=0.47\textwidth]{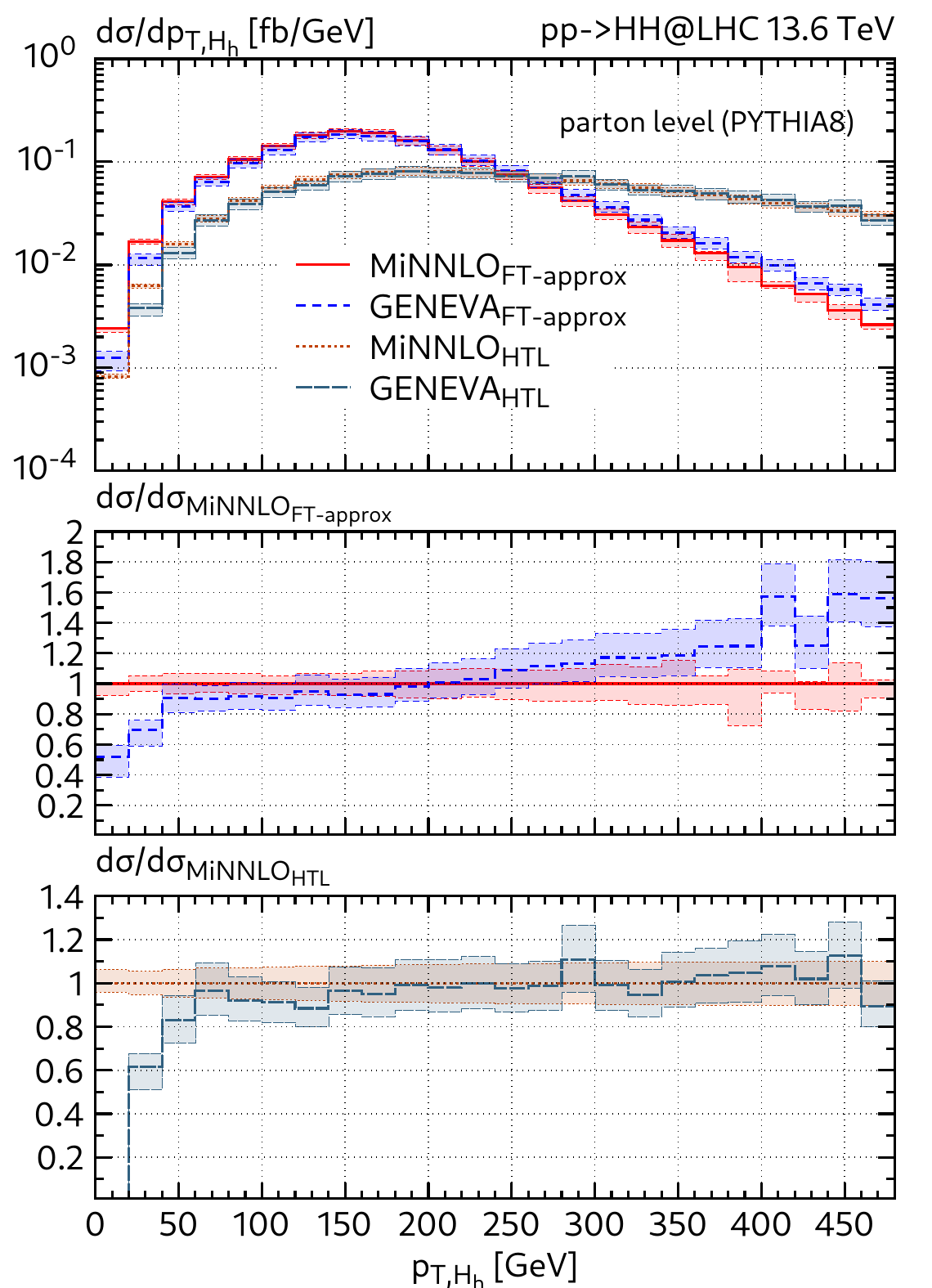}
    \includegraphics[width=0.47\textwidth]{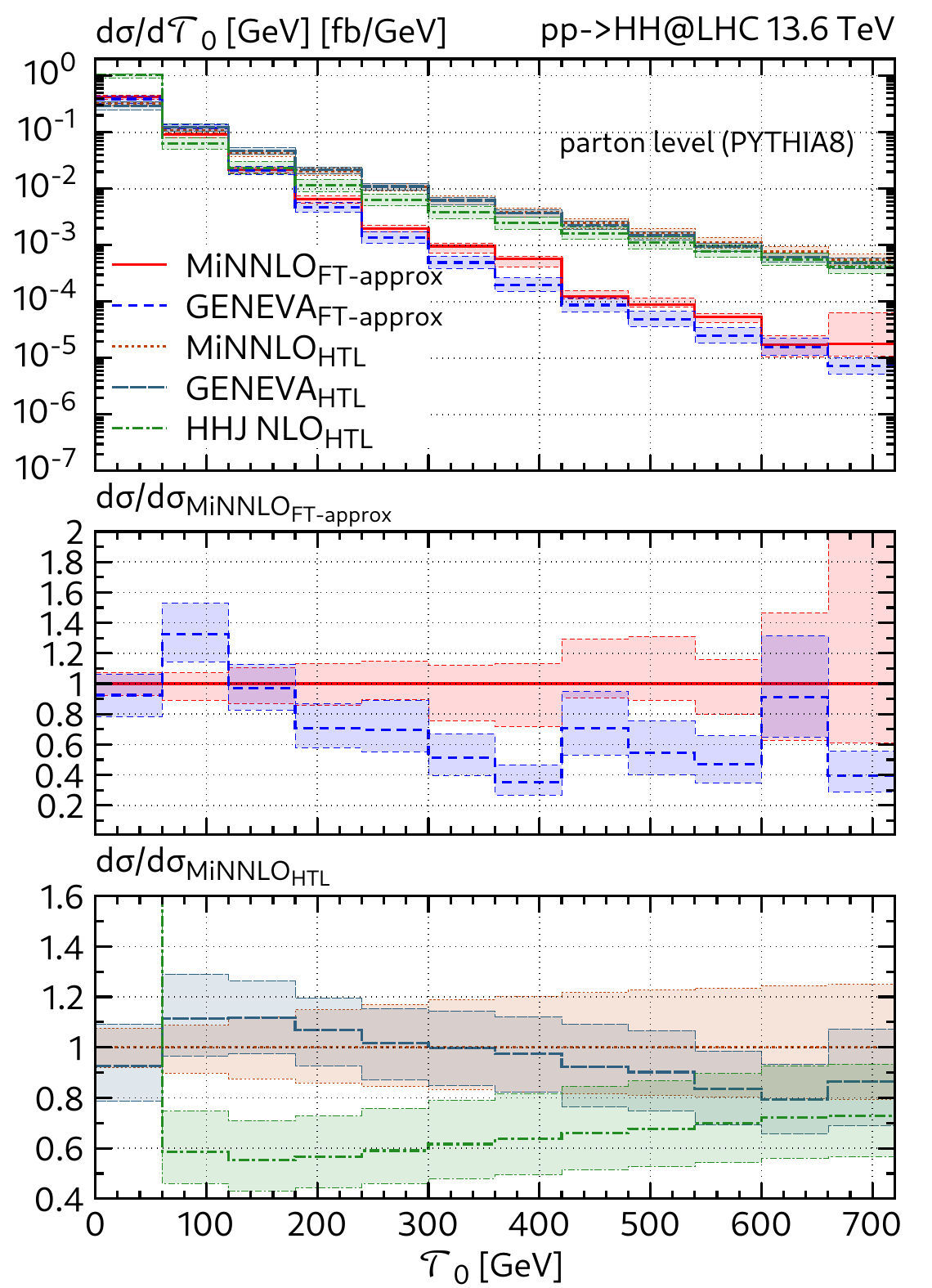}
    \includegraphics[width=0.47\textwidth]{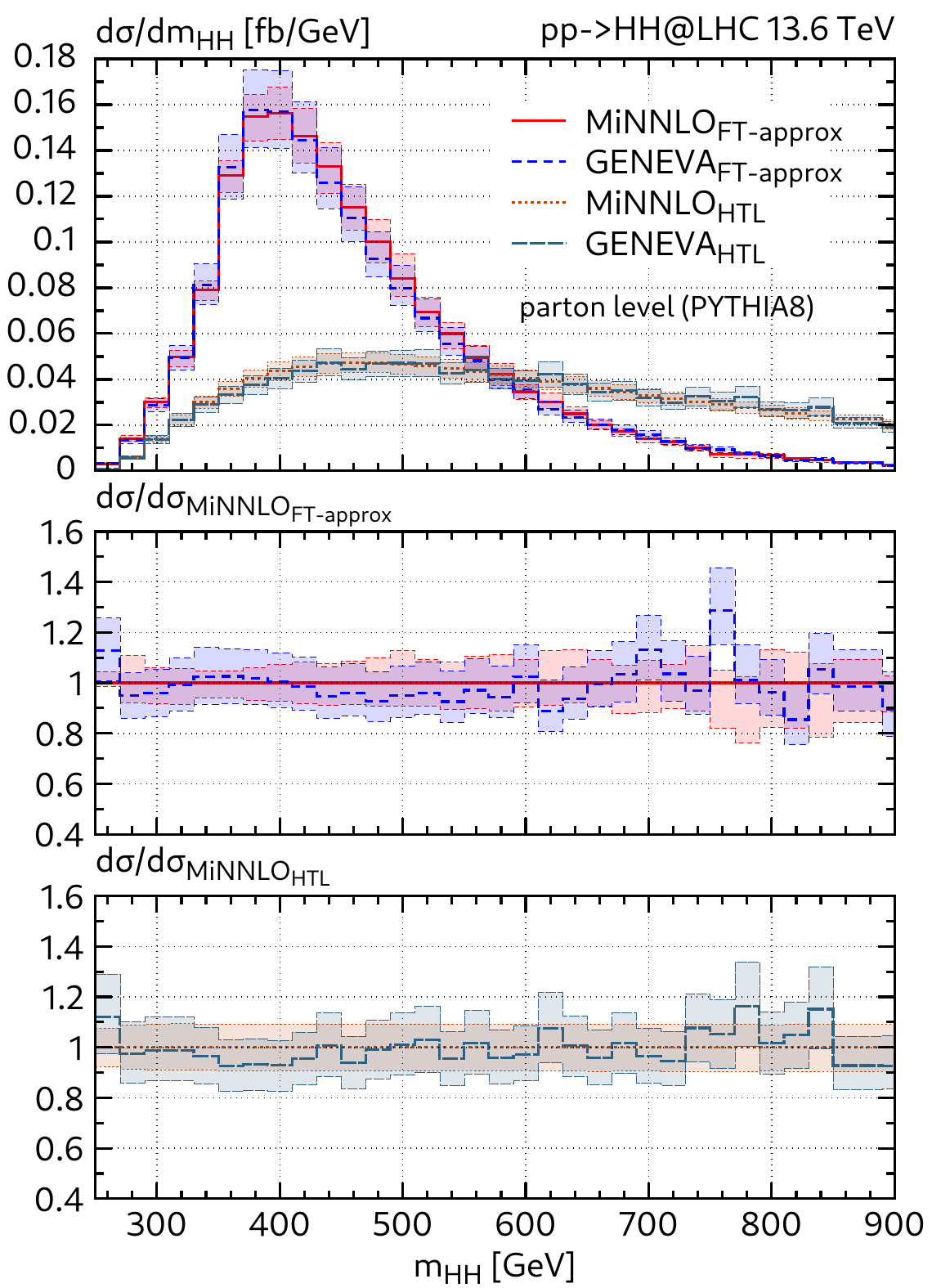}
    \caption{Comparison of \minnlo{} and \textsc{Geneva} predictions in
    the HTL and FT-approx setups. We show
    distributions in the transverse momentum of the Higgs pair
    (top-left), the transverse momentum of the hardest Higgs boson
    (top-right), the zero-jettiness variable $\mathcal{T}_0$ (bottom-left),
    and the invariant mass of the Higgs pair  (bottom-right). For $p_{T,HH}$ and $\mathcal{T}_0$ we also show the result in the HTL obtained from a \textsc{Powheg} $HH$J run up to stage 2.}
    \label{fig:comparison_geneva}
\end{figure}
\afterpage{\clearpage}

In \fig{fig:comparison_geneva} we compare the \minnlo{} and
\textsc{Geneva} predictions for the distributions in
$p_{\textrm{T},HH}$, $p_{\textrm{T},H_h}$, $m_{HH}$ and the
zero-jettiness variable $\mathcal{T}_0$, defined, as in \citere{Alioli:2025xcu}, as 
\be
\mathcal{T}_0 = \sum_{i=1}^n p_{\textrm{T},i}
e^{-|y_i-y_{HH}|}\,,
\label{eq:Tau_0}
\ee
where the sum runs over all final-state QCD partons.
We focus on the comparison in the HTL and using FT-approx,
while the results in the \borni{} approximation show a very similar
behaviour.

We first recall that both the $p_{\textrm{T},HH}$ and
$\mathcal{T}_0$ distributions are effectively only NLO accurate in
their tails, since these observables require recoil against QCD
radiation. By contrast, the low- and intermediate-$p_{\textrm{T},HH}$
and $\mathcal{T}_0$ regions are particularly sensitive to resummation
and matching effects. For these two observables, we also show the NLO prediction obtained by running the $HH$J \textsc{Powheg} code up to stage 2. The fixed-order result becomes unreliable in the low-$p_{\rm T,HH}$ or low $\mathcal{T}_{0}$ region, where large logarithmic corrections spoil the perturbative expansion and resummation is required to obtain reliable predictions. For both observables, this region extends up to approximately $100\,\GeV$. Above this scale, the fixed-order prediction is well behaved, indicating that logarithmically enhanced contributions are under control and resummation effects become less important.
Similarly, the tail of the
$p_{\textrm{T},H_h}$ distribution receives sizeable contributions from
configurations involving hard QCD radiation. It is therefore not
surprising that these observables exhibit the largest differences
between the two calculations.

The most sizeable discrepancies are observed in the
$p_{\textrm{T},HH}$ distribution. In the HTL approximation, the two
predictions exhibit noticeably different shapes, but remain largely
compatible within scale uncertainties across the considered kinematic
range. Once finite top-quark mass effects are included, however, the
differences become substantially larger. In the high-$p_{\textrm{T},HH}$
region, where the cross section is dominated by hard QCD radiation, the
\minnlo{} prediction remains close to the fixed-order result, as shown
in \sct{sec:matrix}, while the \textsc{Geneva} prediction is
significantly larger. This behaviour may be related to the fact that,
within the \textsc{Geneva} framework, $\mathcal{T}_0$ resummation
effects can affect the transverse-momentum spectrum over a broad
kinematic range.

For the $p_{\textrm{T},H_h}$ distribution, the agreement is generally
good in the HTL approximation, whereas noticeable deviations emerge in
the high-transverse-momentum tail once finite top-quark mass effects are
included. These differences are closely related to those observed in the
$p_{\textrm{T},HH}$ distribution due to the strong kinematic correlation
between the two observables, although they are somewhat less pronounced
since $p_{\textrm{T},H_h}$ is less sensitive to the recoil against hard
QCD radiation.

For the $\mathcal{T}_0$ distribution, the \minnlo{} and
\textsc{Geneva} predictions in the HTL approximation agree within
uncertainties, while there 
are some differences in the shape of the two distributions. 
We note that, prior to parton showering, \textsc{Geneva} resums small
$\mathcal{T}_0$ at NNLL$'$ accuracy and that the numerical
impact of the subsequent parton shower remains moderate, even though the
formal accuracy after showering is reduced to LL.\footnote{We note that, unlike other physical observables, $\mathcal{T}_0$ is
subject to enormous hadronisation and MPI effects, which reach up to 300\% in the resummation and matching region.}
 By contrast, the accuracy of the \minnlo{} prediction at small $\mathcal{T}_0$ is entirely
provided by the parton shower and is therefore LL from the outset. In
view of these differences, the overall level of agreement is very
reasonable. Once finite top-quark mass effects are included, however,
the agreement deteriorates, particularly in the tail of the
distribution, while the low-$\mathcal{T}_0$ region, where resummation
effects dominate, remains comparatively stable. 

Both the observed discrepancies and their enhancement once finite
top-quark mass effects are included deserve further investigation. By
contrast, for genuinely NNLO-accurate observables inclusive over QCD
radiation, such as $m_{HH}$, we find good agreement between the two
approaches, both in the HTL and once finite top-quark mass effects are
included.

\section{Results with Higgs decays}
\label{sec:higgsdecays}
We now relax the on-shell treatment of the Higgs bosons and include
their decays in the zero-width approximation through the parton shower. 
We allow for two decays, $H\to\gamma\gamma$ and $H\to b\bar{b}$, imposing fiducial cuts inspired by the experimental analyses. 
In particular, both
$b\bar{b}\gamma\gamma$ ($2b2\gamma$) and $b\bar{b}b\bar{b}$ ($4b$) channels
are studied, following the analyses performed by the ATLAS experiment. The results are shown in FT-approx.


 \subsection[$2b2\gamma$ final state]{\boldmath $2b2\gamma$ final state}

For the $2b2\gamma$ channel, we follow the analysis of
\citere{ATLAS:2025hhd}. 
Since QED effects in
the parton shower are disabled,
 the only final-state photons originate from Higgs-boson
decays and can be used directly to reconstruct one of the two Higgs
bosons. We require the two photons to satisfy the following selection
criteria:
\be
\begin{split}
&\pt(\gamma_1)>35\,\GeV,\qquad \pt(\gamma_2)>25\,\GeV,\qquad 105\,\GeV<m_{\gamma\gamma}<160\,\GeV,\\
&|\eta(\gamma_{1,2})|<2.37,\qquad \frac{\pt(\gamma_1)}{m_{\gamma\gamma}}>0.35,\qquad \frac{\pt(\gamma_2)}{m_{\gamma\gamma}}>0.25.  
\end{split} \label{eq:gamma_sel}
\ee
To reconstruct the other Higgs boson, we require at least two
$b$-tagged anti-$k_\textrm{T}$ jets~\cite{Cacciari:2008gp} with radius
parameter $R=0.4$. Candidate jets are required to satisfy
\be
\pt(j)>25\,\GeV,\qquad |\eta(j)|<2.5\,.
\label{eq:2b_jet_sel}
\ee
If more than two candidate jets are present, we select the pair whose
invariant mass ($m_{bb}$) is closest to the Higgs-boson mass and require
\be
80\,\GeV<m_{bb}<140\,\GeV\,.
\ee
Events containing six or more candidate jets are rejected.

\begin{figure}[t]
    \centering
    \includegraphics[width=0.6\textwidth]{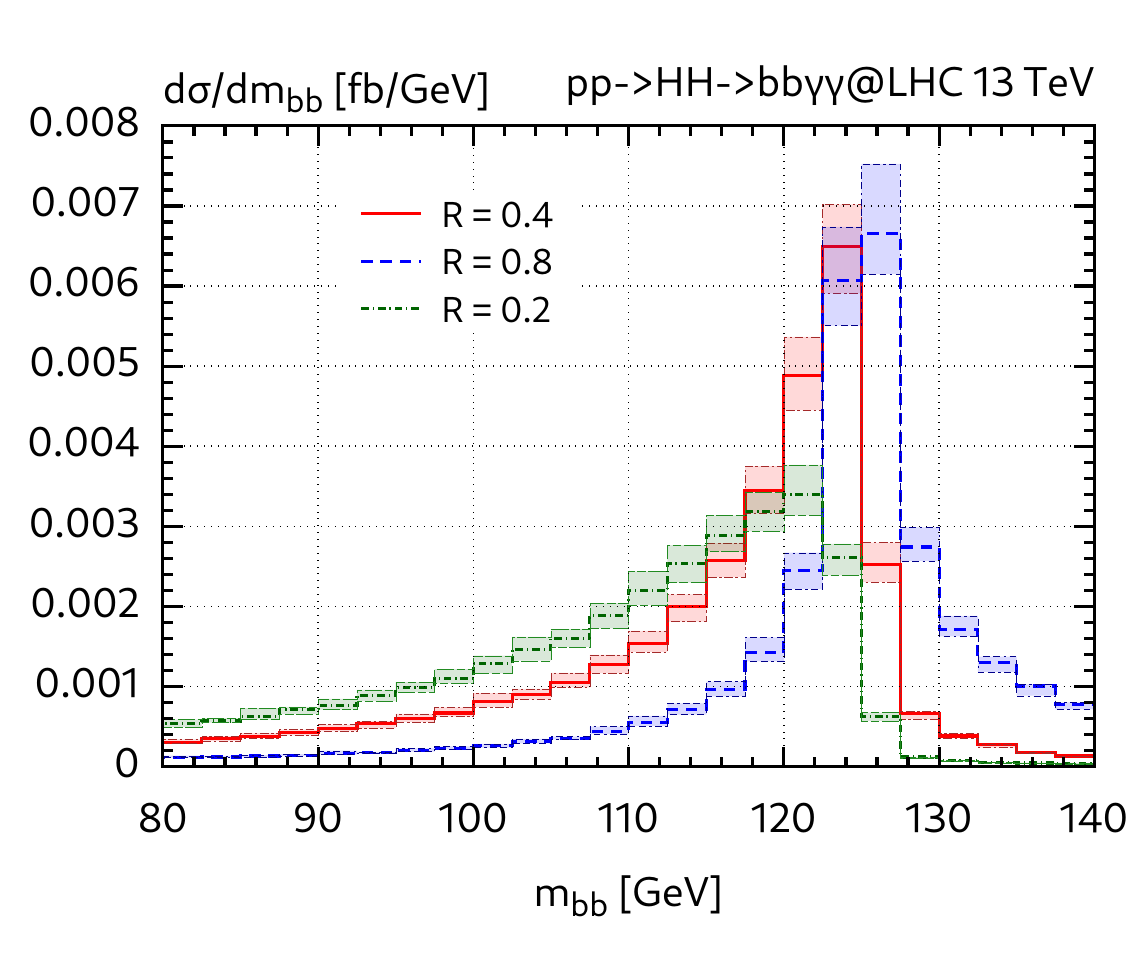}
    \caption{Dijet invariant mass distribution used to reconstruct one of the two Higgs bosons in the $2b2\gamma$ channel. The \minnlo{} prediction is shown in the \FTa{2} approximation.}
    \label{fig:mH1_bbaa}
\end{figure}

In \fig{fig:mH1_bbaa} we show the invariant-mass distribution of the
Higgs boson reconstructed from the two $b$-jets, denoted by $m_{bb}$, as a representative example, for three different values of the jet radius $R$. The second Higgs
boson is reconstructed unambiguously from the diphoton invariant mass
and is otherwise used only to impose the fiducial selection cuts
described above. 
For the standard jet radius $R=0.4$, the $m_{bb}$ distribution exhibits
a relatively broad structure below the Higgs-boson mass, rising towards
a peak around $125\,\GeV$ and falling off rather steeply at larger
invariant masses. Since the underlying $H\to b\bar b$ decay is treated
on shell, this behaviour originates entirely from QCD radiation that is
clustered into the reconstructed $b$-jets and affects the Higgs
reconstruction. 
For larger jet radii, the peak becomes narrower and more symmetric
around the Higgs-boson mass, since final-state radiation from the
bottom quarks is more efficiently clustered into the two $b$-jets. At
the same time, the high-$m_{bb}$ tail becomes more pronounced due to
additional initial-state radiation captured inside the $b$-jets. 
By contrast, for smaller jet radii the distribution exhibits a broad
peak below the Higgs-boson mass, since part of the final-state
radiation is not captured by the reconstructed $b$-jets.


 \subsection[$4b$ final state]{\boldmath $4b$ final state}

For the $4b$ channel, we follow the ATLAS analysis of
\citere{ATLAS:2023qzf}. Jets are reconstructed with the
anti-$k_\textrm{T}$ algorithm using a radius parameter $R=0.4$ and are
classified as central or forward jets according to
\be
\begin{split}
&\text{central jets:}\qquad |\eta|<2.5,\qquad \pt >40\,\GeV,\\
&\text{forward jets:}\qquad 2.5<|\eta|<4.5,\qquad \pt >30\,\GeV.
\end{split}
\label{eq:jets_4b}
\ee
We require at least four $b$-tagged central jets and use the four
hardest ones to reconstruct the two Higgs-boson candidates. For each of
the three possible pairings, we identify the Higgs-boson candidate with
the larger transverse momentum and select the pairing for which this
candidate has the smallest angular separation $\Delta R$ between its
two $b$-jets.

To suppress backgrounds and isolate the gluon-fusion $HH\to4b$ signal,
we closely follow the experimental event selection:
\begin{itemize}
\item
Events in the vector-boson-fusion (VBF) signal region are vetoed. We
search for two non-$b$-tagged jets with the largest invariant mass
$m_{jj}$. If such a pair satisfies
\be
m_{jj}>1\,\TeV,\qquad |\Delta\eta_{jj}|>3,
\ee
and the total transverse momentum of the two VBF jets together with the
four Higgs-candidate jets is smaller than $65\,\GeV$, the event is
classified as VBF-like and rejected.

\item
The pseudorapidity separation of the reconstructed Higgs bosons is
required to satisfy
\be
|\Delta\eta_{HH}|<1.5\,,
\ee
which improves the discrimination between the $HH$ signal and QCD
backgrounds.

\item
To suppress the $t\bar t$ background, we introduce the top-veto
variable
\be
x_{Wt}=
\text{min}\left[
\sqrt{
\left(\frac{m_{jj}-m_W}{0.1\,m_{jj}}\right)^2+
\left(\frac{m_{jjb}-m_t}{0.1\,m_{jjb}}\right)^2
}
\right],
\label{eq:xwt}
\ee
where $m_W=80.4\,\GeV$, $m_t=172.5\,\GeV$, and the factor $0.1$
approximates the experimental dijet mass resolution. Candidate
$W$ bosons are reconstructed from all pairs of central jets with
invariant mass $m_{jj}$, while top-quark candidates with mass
$m_{jjb}$ are obtained by combining the $W$ candidate with any
remaining jet associated with one of the Higgs candidates. Events with
$x_{Wt}<1.5$ are discarded.
\end{itemize}

\begin{figure}[t]
    \centering
    \includegraphics[width=0.47\textwidth]{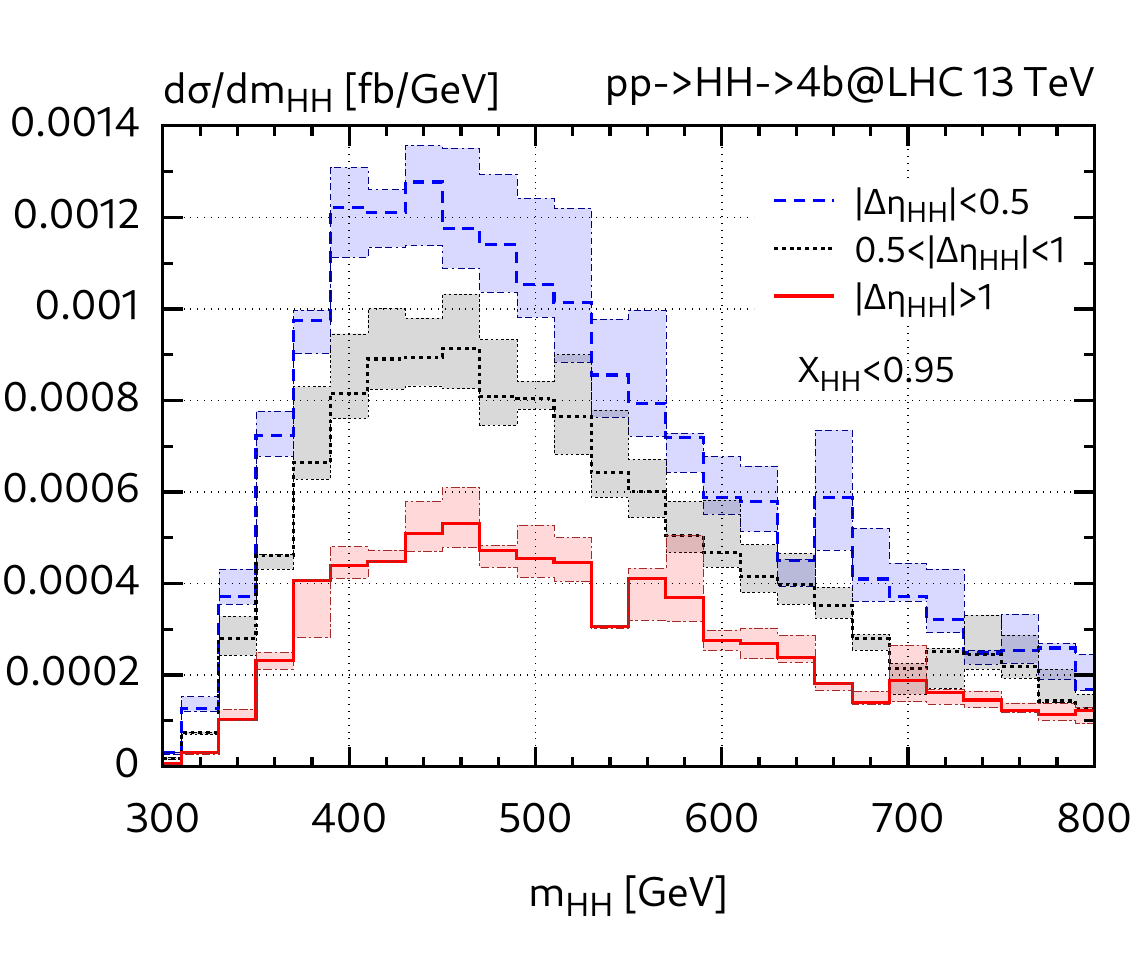}
    \includegraphics[width=0.47\textwidth]{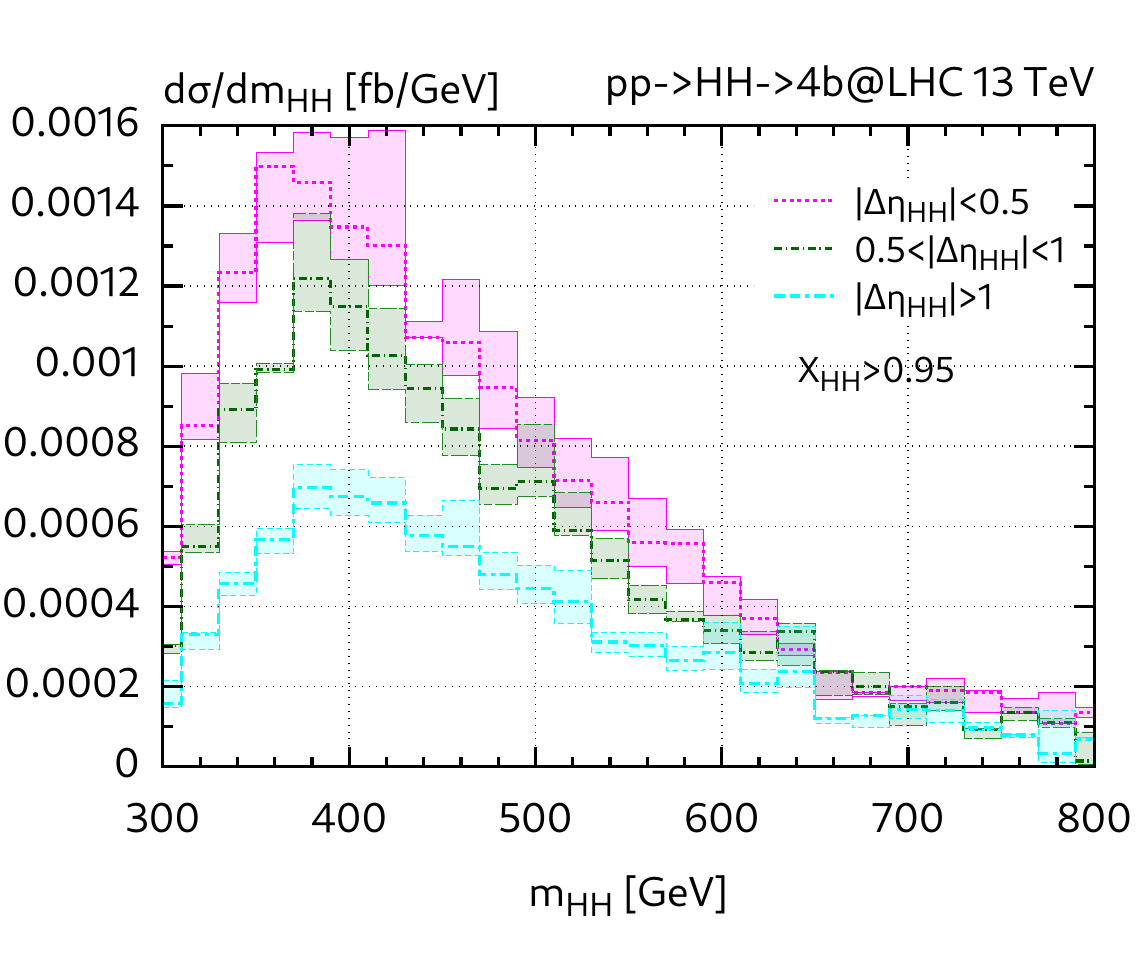}
    \caption{Differential distribution of the Higgs-pair invariant mass
    in the \FTa{2} approximation in the six
    $|\Delta\eta_{HH}|$--$X_{HH}$ regions used by
    ATLAS~\cite{ATLAS:2023qzf}, for $X_{HH}<0.95$ (left) and
    $X_{HH}>0.95$ (right).}
    \label{fig:mHH_4b}
\end{figure}

Following the ATLAS analysis~\cite{ATLAS:2023qzf}, in
\fig{fig:mHH_4b} we show the $m_{HH}$ distribution in six regions of
phase space defined through the pseudorapidity separation of the
reconstructed Higgs bosons $|\Delta\eta_{HH}|$ and the variable
\be
X_{HH}=
\sqrt{
\left(\frac{m_{H_1}-124\,\GeV}{0.1m_{H_1}}\right)^2+
\left(\frac{m_{H_2}-117\,\GeV}{0.1m_{H_2}}\right)^2
},
\label{eq:xhh}
\ee
which quantifies the consistency of the reconstructed dijet systems with
the $HH\to4b$ hypothesis. Here, $m_{H_1}$ and $m_{H_2}$ denote the
invariant masses of the leading and subleading reconstructed
Higgs-boson candidates, respectively. The left and right panels
correspond to the regions with $X_{HH}<0.95$ and $X_{HH}>0.95$,
respectively. 
As expected, the largest rates are found in the most central
$|\Delta\eta_{HH}|$ regions. A visible difference is observed between
the two $X_{HH}$ categories: while the distributions for
$X_{HH}<0.95$ peak around $m_{HH}\simeq 450\,\GeV$, the corresponding
maximum for $X_{HH}>0.95$ is shifted towards smaller invariant
masses, slightly below $m_{HH}\simeq 400\,\GeV$.

\section{Impact of the trilinear Higgs coupling}
\label{sec:trilinear}
We conclude our discussion of phenomenological results by considering
the impact of modifications of the trilinear Higgs coupling, as they
arise in various new-physics scenarios. To this end, we study different
values of the modifier $\kappa_\lambda$, defined through
\be
\lambda_3 = \kappa_\lambda \lambda_3^{\text{SM}}
= \kappa_\lambda \frac{m_H^2}{2v}\,.
\label{eq:lambda3}
\ee
In \tab{tab:sigma_kl} we report total cross sections in both the HTL
and \textrm{FT-approx-}0 approximations\footnote{In this section we employ the \textrm{FT-approx-}0 approximation. While the study could also be performed in the other two FT approximations using
\textsc{ggxy}, the corresponding \textsc{HHgrid} implementation does
not allow for modifications of the trilinear Higgs coupling. Since the
differences among the three FT approximations are small (see
\sct{sec:results_mass}), restricting the analysis to \textrm{FT-approx-}0 is
sufficient for the present study while significantly reducing the
computational cost.} for the values
$\kappa_\lambda=\left\{-1,0,1,2,2.5,4\right\}$, all compatible with the
current bounds reported in \citere{ATLAS:2023qzf},
$-1.7<\kappa_\lambda<6.6$.

\begin{table}[t]
\begin{center}
\begin{tabular}{ | c | c | c | c | c | c | c | c | }
\hline
$\kappa_\lambda$ & $-1$ & $0$ & $1$ (SM) & $2$ & $2.5$ & $4$ \\
\hline
\rule[-1.5ex]{0pt}{5ex} $\sigma_{\textrm{HTL}}$ (fb) & $80.21^{+9.3\%}_{-8.8\%}$ & $49.52^{+8.9\%}_{-8.7\%}$ & $30.08^{+8.1\%}_{-8.5\%}$ & $21.96^{+8\%}_{-8.5\%}$ & $22.13^{+8.6\%}_{-8.6\%}$ & $39.57^{+10\%}_{-9\%}$ \\
\hline
\rule[-1.5ex]{0pt}{5ex} $\sigma_{\textrm{FT-approx-}0}$ (fb) & $128.29^{+7.5\%}_{-8.1\%}$ & $69.82^{+7.3\%}_{-7.2\%}$ & $32.04^{+6.1\%}_{-6.6\%}$ & $15.38^{+4.2\%}_{-6.5\%}$ & $14.74^{+6.4\%}_{-7.3\%}$ & $45.39^{+10.9\%}_{-9.3\%}$ \\
\hline
\end{tabular}
\caption{Total cross sections (in fb)  in the HTL and \textrm{FT-approx-}0 for different $\kappa_\lambda$ values at $13\,\TeV$.\label{tab:sigma_kl}}
\end{center}
\end{table}
\begin{figure}[t]
    \centering
    \includegraphics[width=0.5\textwidth]{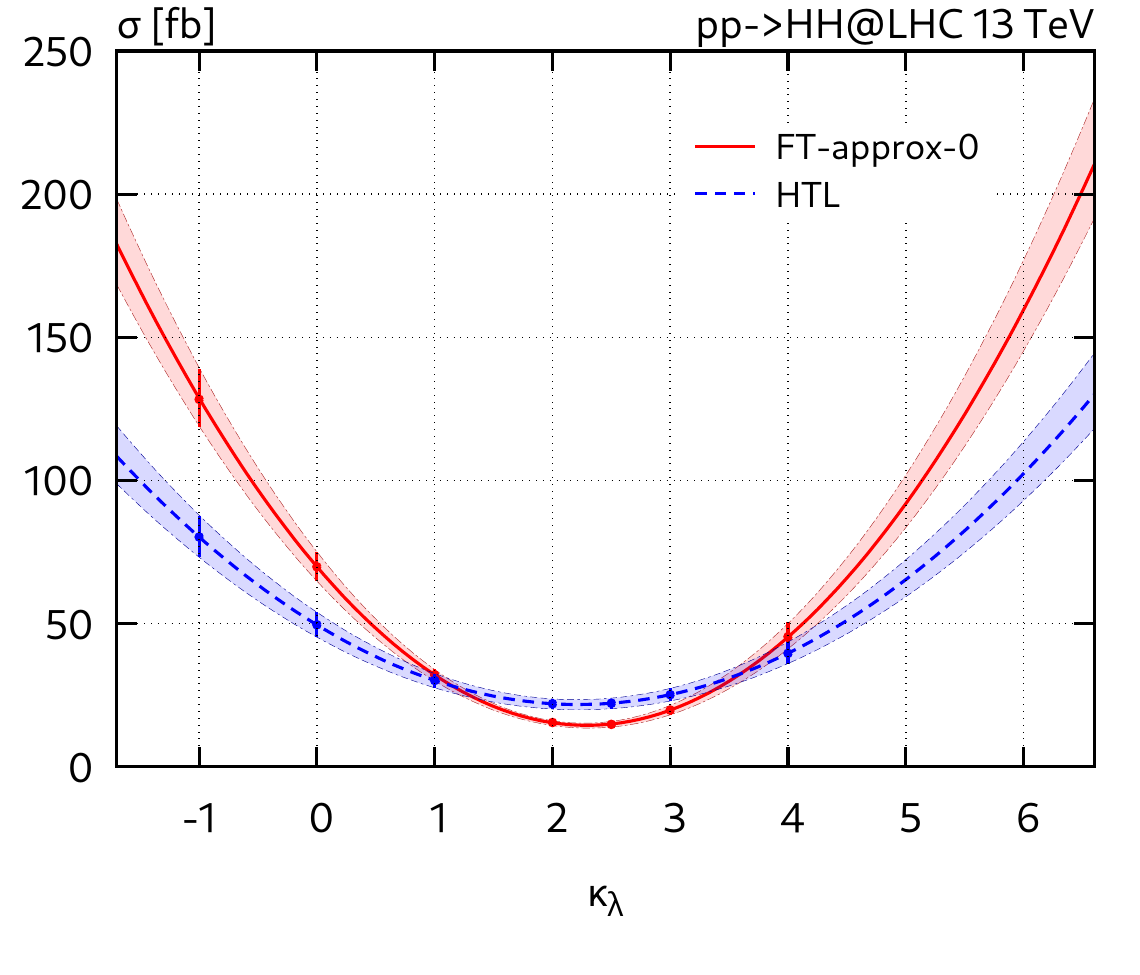}
    \caption{Total cross sections as a function of $\kappa_\lambda$ at
$13\,\TeV$ in the HTL (blue) and \textrm{FT-approx-}0 (red). The
data points correspond to the cross sections reported in
\tab{tab:sigma_kl}, while the curves show their parabolic fits.}
    \label{fig:varying_klambda_fit}
\end{figure}

The cross section receives contributions from the squared box amplitude
(independent of $\kappa_\lambda$), the squared triangle contribution
($\propto \kappa_\lambda^2$), and their interference
($\propto \kappa_\lambda$), which is sizeable and destructive in the SM.
Consequently, the cross section exhibits a characteristic parabolic
behaviour as a function of $\kappa_\lambda$. This behaviour is illustrated
in \fig{fig:varying_klambda_fit}, where we show the cross sections from
\tab{tab:sigma_kl} together with corresponding parabolic fits. 
The minimum occurs around $\kappa_\lambda\simeq 2.2$, where the destructive
interference is maximal, while the $HH$ cross section exceeds the SM
value for $\kappa_\lambda<1$ and $\kappa_\lambda>4$.

It is interesting to observe that the HTL result exhibits a broader
parabola than the corresponding \textrm{FT-approx-}0 result, while the minima of the
two curves are located at the same value of
$\kappa_\lambda$. This provides another clear indication of the importance
of finite top-quark mass effects in Higgs-boson pair production.
Although the HTL and \textrm{FT-approx-}0 predictions are rather close for the SM
value $\kappa_\lambda=1$, they differ substantially for other values of
$\kappa_\lambda$.

\begin{figure}[t]
    \centering
    \includegraphics[width=0.47\textwidth]{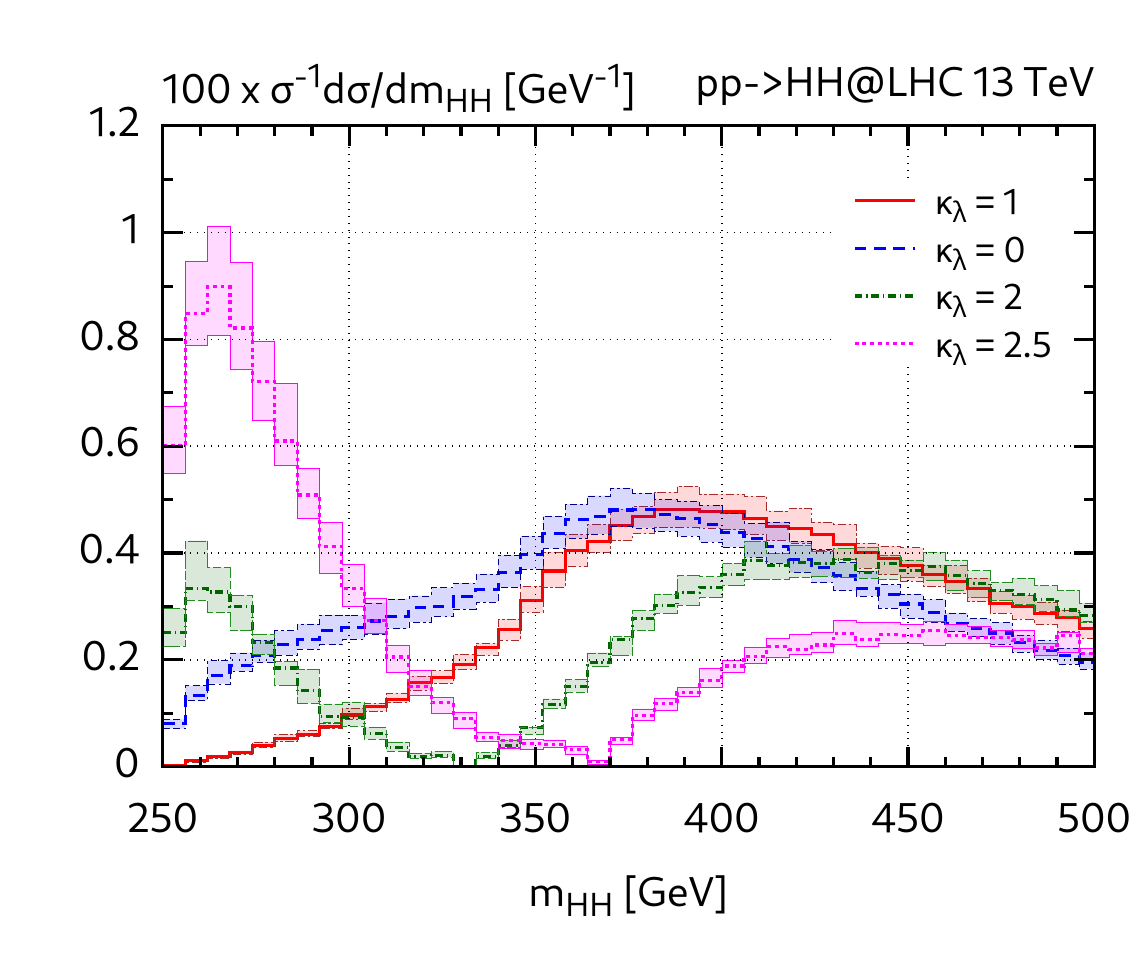}
    \includegraphics[width=0.47\textwidth]{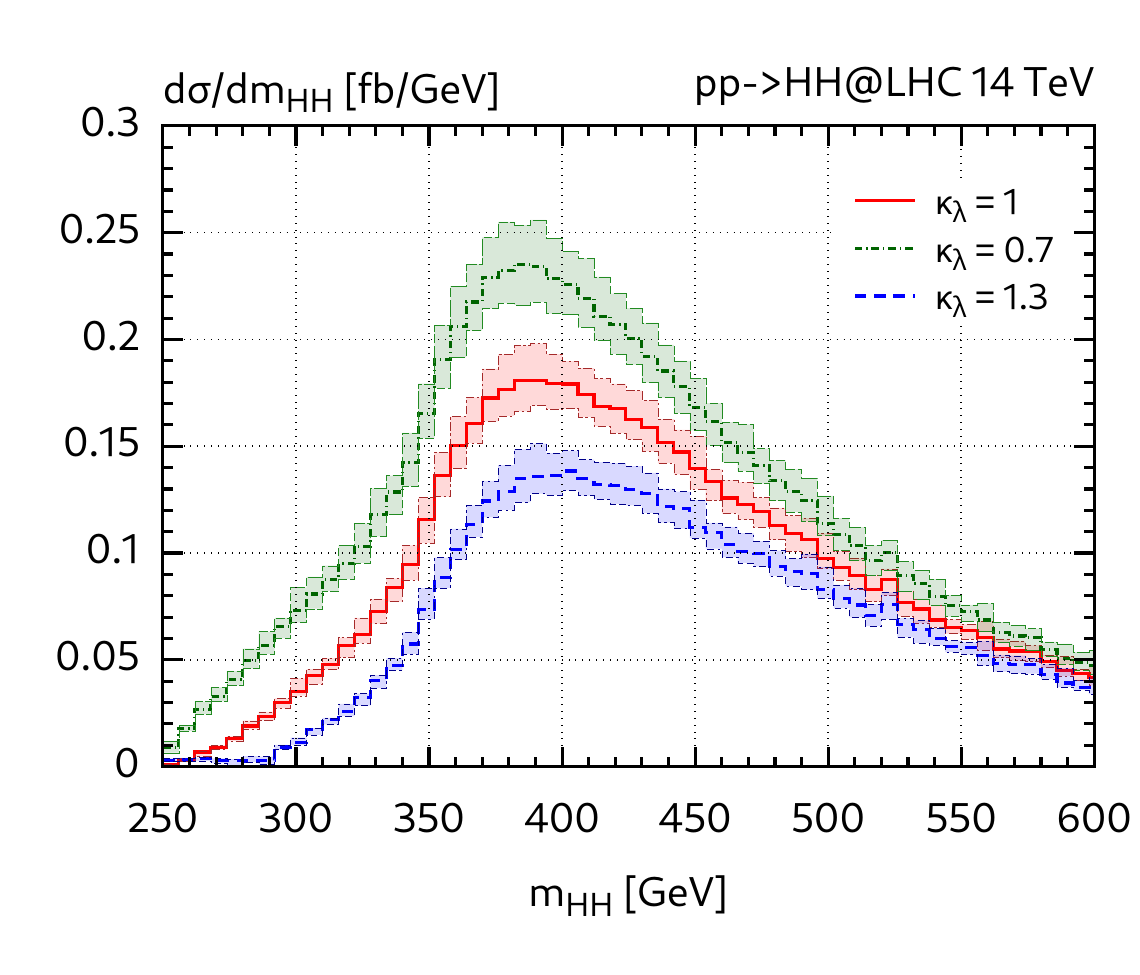}
\caption{Differential distributions in the Higgs-pair invariant mass using \textrm{FT-approx-}0 for different values of $\kappa_\lambda$. Left:
normalised $m_{HH}$ distributions at $13\,\TeV$. Right: absolute
$m_{HH}$ distributions at $14\,\TeV$ for values of $\kappa_\lambda$
compatible with the expected HL-LHC bounds.}
    \label{fig:varying_klambda_mHH}
\end{figure}

We next study the dependence of the $m_{HH}$ distribution on the trilinear Higgs coupling.
In the left panel of \fig{fig:varying_klambda_mHH} we show the  \textrm{FT-approx-}0 predictions of the
$m_{HH}$ distributions for different values of
$\kappa_\lambda$ at $13\,\TeV$. In order to highlight the shape differences, each
distribution is normalised to its corresponding total cross section.
The substantial differences in terms of shape can be understood 
from the interplay of the box,
triangle, and interference contributions. The case $\kappa_\lambda=0$
corresponds to the pure box contribution, whose distribution exhibits a
maximum around $m_{HH}\sim 380\,\GeV$. By contrast, both the triangle
contribution and its interference with the box are enhanced close to
threshold due to the Higgs propagator,
$(\hat{s}-m_H^2)^{-1}$, and therefore predominantly affect the
low-$m_{HH}$ region, while becoming increasingly suppressed at larger
invariant masses. In the SM, the sizeable destructive interference
between box and triangle contributions leads to a suppression of the
cross section close to threshold. Increasing $\kappa_\lambda$ enhances the
triangle contribution quadratically, whereas the interference term grows
only linearly. As a consequence, the low-$m_{HH}$ region becomes
progressively enhanced and eventually develops a pronounced peak as the
triangle contribution starts to dominate over the destructive
interference.

In the right panel of \fig{fig:varying_klambda_mHH} we show the same
distribution at $14\,\TeV$ for the values
$\kappa_\lambda=\left\{0.7,1,1.3\right\}$, corresponding approximately to
the expected HL-LHC sensitivity~\cite{ATLAS:2025eii}. In this case the
shape differences are considerably smaller and we therefore present the
absolute distributions without normalisation.

\section{Conclusions}
\label{sec:conclusions}

In this work we have presented an NNLO+PS event generator for Higgs-boson pair production in gluon fusion within the \minnlo{} framework. Finite top-quark mass effects are incorporated through approximations based on the exact NLO QCD result, making use of the available calculations of the two-loop amplitudes in the full theory. The implementation has been realised within the \textsc{Powheg-Box-Res} framework and is made publicly available together with this publication.

In contrast to single-Higgs production, top-quark mass effects are indispensable for a reliable description of $HH$ production, while the computation of the complete NNLO corrections in the full theory is currently beyond reach. 
We have therefore introduced several approximations to account for top-quark mass effects beyond NLO. Besides the heavy-top limit and the Born-improved approximation, we have considered three different full-theory-inspired setups, which differ in their treatment of the virtual amplitudes, while retaining the exact description of the real radiation. Among them is the standard FT-approx approach. This allows us to assess the residual uncertainty associated with the propagation of mass effects beyond NLO, where exact full-theory information is currently available.

We have presented phenomenological results for stable Higgs bosons at the LHC and compared our predictions to fixed-order NNLO QCD calculations obtained with \textsc{Matrix}. We found very good agreement for inclusive and differential observables, thereby validating our implementation. 

We have further compared our results to recent NNLO+PS predictions obtained within the \textsc{Geneva} framework. 
We find good agreement for observables inclusive over QCD radiation, such as $m_{HH}$. By contrast, transverse-momentum distributions sensitive to QCD radiation exhibit sizeable differences, which increase towards large transverse momenta. 
In this region, which is typically dominated by hard QCD radiation, the \minnlo{} prediction agrees well with the fixed-order result, while the \textsc{Geneva} result is significantly larger. This 
might reflect that the \textsc{Geneva} approach distributes $\mathcal{T}_0$ resummation effects over a broad transverse-momentum range. 
These discrepancies are more pronounced when both calculations include mass effects, compared to the HTL case.
Since mass effects are included using the same approximation in the two calculations, this behaviour is not expected and deserves further investigation trough a dedicated study.

Our results confirm that finite top-quark mass effects are essential for a reliable description of Higgs-pair production. While the heavy-top limit fails to describe both the normalisation and the shape of key observables, the Born-improved approximation captures the main features of inclusive observables, but becomes insufficient for distributions that are directly sensitive to hard QCD radiation.
 By contrast, the different full-theory-inspired approximations considered in this work lead to rather similar predictions, with differences that are generally covered by the corresponding uncertainty bands. This indicates that the remaining ambiguity in the treatment of mass effects beyond NLO is moderate for the observables studied here. 
 It will be interesting to assess these approximations once additional
full-theory ingredients become available, which can then be
straightforwardly incorporated into our \minnlo{} generator.

We have also extended our analysis beyond stable Higgs bosons by considering decays into the $b\bar b\gamma\gamma$ and $b\bar bb\bar b$ final states, implementing fiducial selections inspired by recent ATLAS analyses. In addition, we have illustrated the impact of modifications of the trilinear Higgs coupling on total rates and differential distributions. These results highlight the flexibility of the generator and demonstrate that it can be used directly in realistic phenomenological studies and future experimental applications.

The calculation presented here constitutes an important development in precision simulations of Higgs-pair production at the LHC. On the one hand, it provides a fully differential NNLO+PS description in a public and efficient Monte Carlo implementation. On the other hand, it offers a framework in which future improvements can be incorporated straightforwardly, in particular more accurate information on the virtual amplitudes in the full theory. Such improvements will be important to further reduce the theoretical uncertainties in Higgs-pair production and to strengthen the interpretation of future measurements of the Higgs self-coupling.

\vspace{0.5cm}

\noindent {\bf Public release of the code.}
Together with this manuscript, we release the code {\tt POWHEG-BOX-RES/HHJ}, referred to as the \texttt{HH-MiNNLO} event generator, making it publicly available for experimental analyses and phenomenological studies.
The code can be downloaded by following the instructions on the \POWHEGBOX{} website: \url{http://powhegbox.mib.infn.it}.
To obtain the implementation, navigate to the \texttt{POWHEG-BOX-RES} git directory and execute
\lstset{basicstyle=\small, frame=none}
{\tt
\begin{lstlisting}[language=bash]
  $ git submodule update --remote --init HHJ
\end{lstlisting}
}
The \texttt{HH-MiNNLO} generator provides NLO+PS predictions for $HHJ$ observables and NNLO+PS simulations for $HH$ production based on the combined two-loop approximation described in this work. 
All required dependencies are documented in the manual located at 

\noindent \texttt{HHJ/Docs/manual\_HH-MiNNLO.pdf}.

\section*{Acknowledgments}
We would like to thank Simone Alioli, Giulia Marinelli and Davide Napoletano for providing their results for the comparison presented in \sct{sec:geneva} and for useful discussions.
We are indebted to Javier Mazzitelli for providing us with the central fixed-order NNLO predictions from {\sc Matrix} from \citeres{Grazzini:2018bsd}. 
We have used the Max Planck Computing and Data Facility (MPCDF) in Garching to carry out all simulations presented here.
We thank Florian König and Jakob Linder for providing the container installation to run the code on the MPCDF cluster.  FG is indebted to Christian Biello for the help with the code in the first stage of this work and for useful discussions.

\newpage
\appendix

\section{Hard functions for $gg\to HH$}
\label{app:H1_H2_ggHH}
In the HTL we implement the analytic expressions reported in \citeres{deFlorian:2013jea,deFlorian:2016uhr}. In particular, defining
\be
C_{\text{LO}} = \frac{3m_H^2}{\hat{s}-m_H^2+im_H\Gamma_H}-1, \label{eq:clo}
\ee
where $\hat{s}$ is the partonic centre-of-mass energy, the Born amplitude reads 
\be
\langle \mathcal{R}_{HH}^{{(0)}} |\mathcal{R}_{HH}^{{(0)}}\rangle_{\rm HTL}(\mu_R) = \left(\frac{\alpha_s(\mu_R)}{2\pi}\right)^2\frac{\hat{s}^2G_F^2}{144}|C_{\text{LO}}|^2. \label{eq:born_hh}
\ee
We build the virtual amplitudes through the hard functions $H^{(1)}$ and $H^{(2)}$, defined as 
\be
H^{(1)} =	\frac{2\,{\rm Re}\langle \mathcal{R}_{HH}^{{(0)}} |\mathcal{R}_{HH}^{{(1)}}\rangle}{\langle \mathcal{R}_{HH}^{{(0)}} |\mathcal{R}_{HH}^{{(0)}}\rangle}, \qquad H^{(2)} =	\frac{2\,{\rm Re}\langle \mathcal{R}_{HH}^{{(0)}} |\mathcal{R}_{HH}^{{(2)}}\rangle+\langle \mathcal{R}_{HH}^{{(1)}} |\mathcal{R}_{HH}^{{(1)}}\rangle}{\langle \mathcal{R}_{HH}^{{(0)}} |\mathcal{R}_{HH}^{{(0)}}\rangle}.
\ee

For $H^{(1)}$ we need to evaluate the virtual amplitude with renormalisation scale set to $\hat{s}$ and convert the resulting hard function in the \minnlo{} scheme according to
\be
H^{(1)} = 2\left(\frac{2\,{\rm Re}\langle \mathcal{R}_{HH}^{{(0)}} |\mathcal{R}_{HH}^{{(1)}}\rangle(\mu_R=\sqrt{\hat{s}})}{\langle \mathcal{R}_{HH}^{{(0)}} |\mathcal{R}_{HH}^{{(0)}}\rangle(\mu_R=\sqrt{\hat{s}})}+\frac{\pi^2}{4}\right). \label{eq:shift_h1}
\ee
In the HTL, starting from the analytic formulas reported in \citere{deFlorian:2013jea}, we obtain 
\be
H^{(1)}_{\rm HTL}= 11+\frac{7}{2}\pi^2+\frac{4}{3}\frac{\text{Re}(C_{\text{LO}})}{|C_{\text{LO}}|^2}, \label{eq:h1_minnlo_htl}
\ee
while for the FT result (used in the FT-approx and FT-approx-2 approximations) we compute the virtual amplitude with \textsc{ggxy} and apply \eqn{eq:shift_h1}. We then multiply $H^{(1)}$ by the Born amplitude in the HTL or in FT according to the approximation
\begin{align}
\begin{split}\label{h1tovirtual}
2\,{\rm Re}\langle \mathcal{R}_{HH}^{{(0)}} |\mathcal{R}_{HH}^{{(1)}}\rangle_{\rm HTL} &= H^{(1)}_{\rm HTL}\cdot \langle \mathcal{R}_{HH}^{{(0)}} |\mathcal{R}_{HH}^{{(0)}}\rangle_{\rm HTL} \,,\\
2\,{\rm Re}\langle \mathcal{R}_{HH}^{{(0)}} |\mathcal{R}_{HH}^{{(1)}}\rangle_{\textrm{Born-I/FT-approx-}0} &= H^{(1)}_{\rm HTL}\cdot \langle \mathcal{R}_{HH}^{{(0)}} |\mathcal{R}_{HH}^{{(0)}}\rangle_{\rm FT} \,,\\
2\,{\rm Re}\langle \mathcal{R}_{HH}^{{(0)}} |\mathcal{R}_{HH}^{{(1)}}\rangle_{\textrm{FT-approx/FT-approx-}2} &= H^{(1)}_{\rm FT}\cdot \langle \mathcal{R}_{HH}^{{(0)}} |\mathcal{R}_{HH}^{{(0)}}\rangle_{\rm FT} \,.
\end{split}
\end{align}  

For $H^{(2)}$, we use the procedure described in \citere{Becher:2009qa}: we recover the expression for $H^{(1)}$, and we get 
\be
\begin{split}
H^{(2)}_{\rm HTL} &= \frac{37}{8}\pi^4+\frac{1679}{24}\pi^2-\frac{499}{6}\zeta(3)+\frac{5359}{54}+\frac{137}{6}\log\left(\frac{\hat{s}}{m_t^2}\right)\\
& +\frac{(14C_A\pi^2+9\mathcal{R}_2)\text{Re}C_{\text{LO}}+9\mathcal{I}_2\text{Im}C_{\text{LO}}+9\mathcal{V}_2}{9|C_{\text{LO}}|^2},
\end{split} \label{eq:h2_minnlo}
\ee
where the functions $\mathcal{R}_2$, $\mathcal{I}_2$ and $\mathcal{V}_2$ are given by
\be
\begin{split}
\mathcal{R}_2 &= -\left(1+\frac{2m_H^4}{\hat{s}^2}\right)\left\{-\frac{4}{3}\pi^2+2\text{Li}_2\left(1-\frac{2m_H^4}{\hat{t}\hat{u}}\right)+4\text{Li}_2\left(\frac{m_H^2}{\hat{t}}\right)+4\text{Li}_2\left(\frac{m_H^2}{\hat{u}}\right)\right.\\
&\left.+4\log\left(1-\frac{m_H^2}{\hat{t}}\right)\log\left(-\frac{m_H^2}{\hat{t}}\right)+4\log\left(1-\frac{m_H^2}{\hat{u}}\right)\log\left(-\frac{m_H^2}{\hat{u}}\right)-\log^2\left(\frac{\hat{t}}{\hat{u}}\right)\right\}\\
&+\frac{4m_H^2}{\hat{s}}+\frac{209}{9}-\frac{164}{27}n_f-\frac{33-2n_f}{9}\log\left(\frac{\hat{t}\hat{u}}{\hat{s}^2}\right),
\end{split} \label{eq:r2}
\ee
\be
\mathcal{I}_2 = 4\pi\left(1+\frac{2m_H^4}{\hat{s}^2}\right)\log\left(\frac{(m_H^2-\hat{t})(m_H^2-\hat{u})}{\hat{t}\hat{u}}\right), \label{eq:i2}
\ee
\be
\mathcal{V}_2 = \frac{1}{(3\hat{s}\hat{t}\hat{u})^2}\left[m_H^8(\hat{t}+\hat{u})^2-2m_H^4\hat{t}\hat{u}(\hat{t}+\hat{u})^2+\hat{t}^2\hat{u}^2\left(4\hat{s}^2+(\hat{t}+\hat{u})^2\right)\right]. \label{eq:v2}
\ee
We finally get the NNLO contributions to $gg\to HH$ as
\begin{align}
\begin{split}\label{h2tovirtual}
&\left(2\,{\rm Re}\langle \mathcal{R}_{HH}^{{(0)}} |\mathcal{R}_{HH}^{{(2)}}\rangle+\langle \mathcal{R}_{HH}^{{(1)}} |\mathcal{R}_{HH}^{{(1)}}\rangle\right)_{\rm HTL} = H^{(2)}_{\rm HTL}\cdot \langle \mathcal{R}_{HH}^{{(0)}} |\mathcal{R}_{HH}^{{(0)}}\rangle_{\rm HTL} \,,\\
&\left(2\,{\rm Re}\langle \mathcal{R}_{HH}^{{(0)}} |\mathcal{R}_{HH}^{{(2)}}\rangle+\langle \mathcal{R}_{HH}^{{(1)}} |\mathcal{R}_{HH}^{{(1)}}\rangle\right)_{\textrm{Born-I/FT-approx-}0/\textrm{FT-approx}}= H^{(2)}_{\rm HTL}\cdot \langle \mathcal{R}_{HH}^{{(0)}} |\mathcal{R}_{HH}^{{(0)}}\rangle_{\rm FT} \,,\\
&\left(2\,{\rm Re}\langle \mathcal{R}_{HH}^{{(0)}} |\mathcal{R}_{HH}^{{(2)}}\rangle+\langle \mathcal{R}_{HH}^{{(1)}} |\mathcal{R}_{HH}^{{(1)}}\rangle\right)_{\textrm{FT-approx-}3}= H^{(2)}_{\rm HTL}\cdot\frac{H^{(1)}_{\rm FT}}{H^{(1)}_{\rm HTL}}\cdot \langle \mathcal{R}_{HH}^{{(0)}} |\mathcal{R}_{HH}^{{(0)}}\rangle_{\rm FT} \,.
\end{split}
\end{align}  

\vspace{1 cm}
\bibliographystyle{JHEP}
\bibliography{biblio}

\providecommand{\href}[2]{#2}\begingroup\raggedright\begin{thebibliography}{10}

\bibitem{Aad:2012tfa}
{\scshape ATLAS} collaboration, G.~Aad et~al., \emph{{Observation of a new
  particle in the search for the Standard Model Higgs boson with the ATLAS
  detector at the LHC}},
  \href{https://doi.org/10.1016/j.physletb.2012.08.020}{\emph{Phys. Lett. B}
  {\bfseries 716} (2012) 1--29},
  [\href{https://arxiv.org/abs/1207.7214}{{\ttfamily 1207.7214}}].

\bibitem{Chatrchyan:2012xdj}
{\scshape CMS} collaboration, S.~Chatrchyan et~al., \emph{{Observation of a new
  boson at a mass of 125 GeV with the CMS experiment at the LHC}},
  \href{https://doi.org/10.1016/j.physletb.2012.08.021}{\emph{Phys. Lett. B}
  {\bfseries 716} (2012) 30--61},
  [\href{https://arxiv.org/abs/1207.7235}{{\ttfamily 1207.7235}}].

\bibitem{Czakon:2021yub}
M.~Czakon, R.~V. Harlander, J.~Klappert and M.~Niggetiedt, \emph{{Exact
  Top-Quark Mass Dependence in Hadronic Higgs Production}},
  \href{https://doi.org/10.1103/PhysRevLett.127.162002}{\emph{Phys. Rev. Lett.}
  {\bfseries 127} (2021) 162002},
  [\href{https://arxiv.org/abs/2105.04436}{{\ttfamily 2105.04436}}]. [Erratum:
  Phys.Rev.Lett. 131, 179901 (2023)].

\bibitem{Niggetiedt:2024nmp}
M.~Niggetiedt and M.~Wiesemann, \emph{{Higgs-boson production in the full
  theory at NNLO+PS}},
  \href{https://doi.org/10.1016/j.physletb.2024.139043}{\emph{Phys. Lett. B}
  {\bfseries 858} (2024) 139043},
  [\href{https://arxiv.org/abs/2407.01354}{{\ttfamily 2407.01354}}].

\bibitem{Glover:1987nx}
E.~W.~N. Glover and J.~J. van~der Bij, \emph{{Higgs Boson Pair Production via
  Gluon Fusion}},
  \href{https://doi.org/10.1016/0550-3213(88)90083-1}{\emph{Nucl. Phys. B}
  {\bfseries 309} (1988) 282--294}.

\bibitem{Plehn:1996wb}
T.~Plehn, M.~Spira and P.~M. Zerwas, \emph{{Pair production of neutral Higgs
  particles in gluon-gluon collisions}},
  \href{https://doi.org/10.1016/0550-3213(96)00418-X}{\emph{Nucl. Phys. B}
  {\bfseries 479} (1996) 46--64},
  [\href{https://arxiv.org/abs/hep-ph/9603205}{{\ttfamily hep-ph/9603205}}].
  [Erratum: Nucl. Phys. B 531 (1998) 655--655].

\bibitem{Dawson:1998py}
S.~Dawson, S.~Dittmaier and M.~Spira, \emph{{Neutral Higgs boson pair
  production at hadron colliders: QCD corrections}},
  \href{https://doi.org/10.1103/PhysRevD.58.115012}{\emph{Phys. Rev. D}
  {\bfseries 58} (1998) 115012},
  [\href{https://arxiv.org/abs/hep-ph/9805244}{{\ttfamily hep-ph/9805244}}].

\bibitem{Borowka:2016ehy}
S.~Borowka, N.~Greiner, G.~Heinrich, S.~P. Jones, M.~Kerner, J.~Schlenk et~al.,
  \emph{{Higgs Boson Pair Production in Gluon Fusion at Next-to-Leading Order
  with Full Top-Quark Mass Dependence}},
  \href{https://doi.org/10.1103/PhysRevLett.117.012001}{\emph{Phys. Rev. Lett.}
  {\bfseries 117} (2016) 012001},
  [\href{https://arxiv.org/abs/1604.06447}{{\ttfamily 1604.06447}}]. [Erratum:
  Phys. Rev. Lett. 117 (2016) 079901].

\bibitem{Borowka:2016ypz}
S.~Borowka, N.~Greiner, G.~Heinrich, S.~P. Jones, M.~Kerner, J.~Schlenk et~al.,
  \emph{{Full top quark mass dependence in Higgs boson pair production at
  NLO}}, \href{https://doi.org/10.1007/JHEP10(2016)107}{\emph{JHEP} {\bfseries
  10} (2016) 107}, [\href{https://arxiv.org/abs/1608.04798}{{\ttfamily
  1608.04798}}].

\bibitem{deFlorian:2013jea}
D.~de~Florian and J.~Mazzitelli, \emph{{Higgs Boson Pair Production at
  Next-to-Next-to-Leading Order in QCD}},
  \href{https://doi.org/10.1103/PhysRevLett.111.201801}{\emph{Phys. Rev. Lett.}
  {\bfseries 111} (2013) 201801},
  [\href{https://arxiv.org/abs/1309.6594}{{\ttfamily 1309.6594}}].

\bibitem{Grigo:2014jma}
J.~Grigo, K.~Melnikov and M.~Steinhauser, \emph{{Virtual corrections to Higgs
  boson pair production in the large top quark mass limit}},
  \href{https://doi.org/10.1016/j.nuclphysb.2014.09.003}{\emph{Nucl. Phys. B}
  {\bfseries 888} (2014) 17--29},
  [\href{https://arxiv.org/abs/1408.2422}{{\ttfamily 1408.2422}}].

\bibitem{deFlorian:2016uhr}
D.~de~Florian, M.~Grazzini, C.~Hanga, S.~Kallweit, J.~M. Lindert,
  P.~Maierhöfer et~al., \emph{{Differential Higgs Boson Pair Production at
  Next-to-Next-to-Leading Order in QCD}},
  \href{https://doi.org/10.1007/JHEP09(2016)151}{\emph{JHEP} {\bfseries 09}
  (2016) 151}, [\href{https://arxiv.org/abs/1606.09519}{{\ttfamily
  1606.09519}}].

\bibitem{deFlorian:2015moa}
D.~de~Florian and J.~Mazzitelli, \emph{{Higgs pair production at
  next-to-next-to-leading logarithmic accuracy at the LHC}},
  \href{https://doi.org/10.1007/JHEP09(2015)053}{\emph{JHEP} {\bfseries 09}
  (2015) 053}, [\href{https://arxiv.org/abs/1505.07122}{{\ttfamily
  1505.07122}}].

\bibitem{Shao:2013bz}
H.-S. Shao, H.~T. Li, C.~S. Li and J.~Wang, \emph{{Threshold resummation
  effects in Higgs boson pair production at the LHC}},
  \href{https://doi.org/10.1007/JHEP07(2013)169}{\emph{JHEP} {\bfseries 07}
  (2013) 169}, [\href{https://arxiv.org/abs/1301.1245}{{\ttfamily 1301.1245}}].

\bibitem{Davies:2018ood}
J.~Davies, G.~Mishima, M.~Steinhauser and D.~Wellmann, \emph{{Double-Higgs
  boson production in the high-energy limit: planar master integrals}},
  \href{https://doi.org/10.1007/JHEP03(2018)048}{\emph{JHEP} {\bfseries 03}
  (2018) 048}, [\href{https://arxiv.org/abs/1801.09696}{{\ttfamily
  1801.09696}}].

\bibitem{Chen:2021ibm}
L.-B. Chen, H.~T. Li, H.-S. Shao and J.~Wang, \emph{{Higgs boson pair
  production via gluon fusion at N$^3$LO in QCD}},
  \href{https://doi.org/10.1016/j.physletb.2020.135292}{\emph{Phys. Lett. B}
  {\bfseries 803} (2020) 135292},
  [\href{https://arxiv.org/abs/1909.06808}{{\ttfamily 1909.06808}}].

\bibitem{Grazzini:2018bsd}
M.~Grazzini, G.~Heinrich, S.~Jones, S.~Kallweit, M.~Kerner, J.~M. Lindert
  et~al., \emph{{Higgs boson pair production at NNLO with top quark mass
  effects}}, \href{https://doi.org/10.1007/JHEP05(2018)059}{\emph{JHEP}
  {\bfseries 05} (2018) 059},
  [\href{https://arxiv.org/abs/1803.02463}{{\ttfamily 1803.02463}}].

\bibitem{Frederix:2014hta}
R.~Frederix, S.~Frixione, V.~Hirschi, F.~Maltoni, O.~Mattelaer, P.~Torrielli
  et~al., \emph{{Higgs pair production at the LHC with NLO and parton-shower
  effects}}, \href{https://doi.org/10.1016/j.physletb.2014.03.026}{\emph{Phys.
  Lett. B} {\bfseries 732} (2014) 142--149},
  [\href{https://arxiv.org/abs/1401.7340}{{\ttfamily 1401.7340}}].

\bibitem{Alioli:2022dkj}
S.~Alioli, G.~Billis, A.~Broggio, A.~Gavardi, S.~Kallweit, M.~A. Lim et~al.,
  \emph{{Double Higgs production at NNLO interfaced to parton showers in
  GENEVA}}, \href{https://doi.org/10.1007/JHEP06(2023)205}{\emph{JHEP}
  {\bfseries 06} (2023) 205},
  [\href{https://arxiv.org/abs/2212.10489}{{\ttfamily 2212.10489}}].

\bibitem{Alioli:2025xcu}
S.~Alioli, G.~Marinelli and D.~Napoletano, \emph{{NNLO+PS double Higgs boson
  production with top-quark mass corrections in GENEVA}},
  \href{https://doi.org/10.1007/JHEP09(2025)206}{\emph{JHEP} {\bfseries 09}
  (2025) 206}, [\href{https://arxiv.org/abs/2507.08558}{{\ttfamily
  2507.08558}}].

\bibitem{Monni:2019whf}
P.~F. Monni, P.~Nason, E.~Re, M.~Wiesemann and G.~Zanderighi,
  \emph{{MiNNLO$_{\rm PS}$: a new method to match NNLO QCD to parton showers}},
  \href{https://doi.org/10.1007/JHEP05(2020)143}{\emph{JHEP} {\bfseries 05}
  (2020) 143}, [\href{https://arxiv.org/abs/1908.06987}{{\ttfamily
  1908.06987}}].

\bibitem{Jezo:2015aia}
T.~Ježo and P.~Nason, \emph{{On the treatment of resonances in next-to-leading
  order calculations matched to a parton shower}},
  \href{https://doi.org/10.1007/JHEP12(2015)065}{\emph{JHEP} {\bfseries 12}
  (2015) 065}, [\href{https://arxiv.org/abs/1509.09071}{{\ttfamily
  1509.09071}}].

\bibitem{Monni:2020nks}
P.~F. Monni, E.~Re and M.~Wiesemann, \emph{{MiNNLO$_{\rm PS}$: optimizing
  $2\to1$ hadronic processes}},
  \href{https://doi.org/10.1140/epjc/s10052-020-08658-5}{\emph{Eur. Phys. J. C}
  {\bfseries 80} (2020) 1075},
  [\href{https://arxiv.org/abs/2006.04133}{{\ttfamily 2006.04133}}].

\bibitem{Lombardi:2020wju}
D.~Lombardi, M.~Wiesemann and G.~Zanderighi, \emph{{Advancing MiNNLO$_{\rm PS}$
  to diboson processes: $Z\gamma$ production at NNLO+PS}},
  \href{https://doi.org/10.1007/JHEP06(2021)095}{\emph{JHEP} {\bfseries 06}
  (2021) 095}, [\href{https://arxiv.org/abs/2010.10478}{{\ttfamily
  2010.10478}}].

\bibitem{Mazzitelli:2020jio}
J.~Mazzitelli, P.~Nason, E.~Re, M.~Wiesemann and G.~Zanderighi,
  \emph{{Next-to-next-to-leading order event generation for top-quark pair
  production}},
  \href{https://doi.org/10.1103/PhysRevLett.127.062001}{\emph{Phys. Rev. Lett.}
  {\bfseries 127} (2021) 062001},
  [\href{https://arxiv.org/abs/2012.14267}{{\ttfamily 2012.14267}}].

\bibitem{Mazzitelli:2024ura}
J.~Mazzitelli, V.~Sotnikov and M.~Wiesemann, \emph{{Next-to-next-to-leading
  order event generation for $Z$-boson production in association with a
  bottom-quark pair}},  \href{https://arxiv.org/abs/2404.08598}{{\ttfamily
  2404.08598}}.

\bibitem{Lombardi:2021rvg}
D.~Lombardi, M.~Wiesemann and G.~Zanderighi, \emph{{W$^+$W$^-$ production at
  NNLO+PS with MiNNLO$_{\rm PS}$}},
  \href{https://doi.org/10.1007/JHEP11(2021)230}{\emph{JHEP} {\bfseries 11}
  (2021) 230}, [\href{https://arxiv.org/abs/2103.12077}{{\ttfamily
  2103.12077}}].

\bibitem{Mazzitelli:2021mmm}
J.~Mazzitelli, P.~F. Monni, P.~Nason, E.~Re, M.~Wiesemann and G.~Zanderighi,
  \emph{{Top-pair production at the LHC with MINNLO$_{\rm PS}$}},
  \href{https://doi.org/10.1007/JHEP04(2022)079}{\emph{JHEP} {\bfseries 04}
  (2022) 079}, [\href{https://arxiv.org/abs/2112.12135}{{\ttfamily
  2112.12135}}].

\bibitem{Buonocore:2021fnj}
L.~Buonocore, G.~Koole, D.~Lombardi, L.~Rottoli, M.~Wiesemann and
  G.~Zanderighi, \emph{{ZZ production at nNNLO+PS with MiNNLO$_{\rm PS}$}},
  \href{https://doi.org/10.1007/JHEP01(2022)072}{\emph{JHEP} {\bfseries 01}
  (2022) 072}, [\href{https://arxiv.org/abs/2108.05337}{{\ttfamily
  2108.05337}}].

\bibitem{Lombardi:2021wug}
D.~Lombardi, M.~Wiesemann and G.~Zanderighi, \emph{{Anomalous couplings in
  Z{\ensuremath{\gamma}} events at NNLO+PS and improving
  {\ensuremath{\nu}}{\ensuremath{\nu}}{\textasciimacron}{\ensuremath{\gamma}}
  backgrounds in dark-matter searches}},
  \href{https://doi.org/10.1016/j.physletb.2021.136846}{\emph{Phys. Lett. B}
  {\bfseries 824} (2022) 136846},
  [\href{https://arxiv.org/abs/2108.11315}{{\ttfamily 2108.11315}}].

\bibitem{Zanoli:2021iyp}
S.~Zanoli, M.~Chiesa, E.~Re, M.~Wiesemann and G.~Zanderighi,
  \emph{{Next-to-next-to-leading order event generation for VH production with
  H {\textrightarrow}$ b\overline{b} $ decay}},
  \href{https://doi.org/10.1007/JHEP07(2022)008}{\emph{JHEP} {\bfseries 07}
  (2022) 008}, [\href{https://arxiv.org/abs/2112.04168}{{\ttfamily
  2112.04168}}].

\bibitem{Gavardi:2022ixt}
A.~Gavardi, C.~Oleari and E.~Re, \emph{{NNLO+PS Monte Carlo simulation of
  photon pair production with MiNNLO$_{PS}$}},
  \href{https://doi.org/10.1007/JHEP09(2022)061}{\emph{JHEP} {\bfseries 09}
  (2022) 061}, [\href{https://arxiv.org/abs/2204.12602}{{\ttfamily
  2204.12602}}].

\bibitem{Haisch:2022nwz}
U.~Haisch, D.~J. Scott, M.~Wiesemann, G.~Zanderighi and S.~Zanoli, \emph{{NNLO
  event generation for $ pp\to Zh\to
  {\mathrm{\ell}}^{+}{\mathrm{\ell}}^{-}b\overline{b} $ production in the SM
  effective field theory}},
  \href{https://doi.org/10.1007/JHEP07(2022)054}{\emph{JHEP} {\bfseries 07}
  (2022) 054}, [\href{https://arxiv.org/abs/2204.00663}{{\ttfamily
  2204.00663}}].

\bibitem{Lindert:2022qdd}
J.~M. Lindert, D.~Lombardi, M.~Wiesemann, G.~Zanderighi and S.~Zanoli,
  \emph{{W$^{±}$Z production at NNLO QCD and NLO EW matched to parton showers
  with MiNNLO$_{\rm PS}$}},
  \href{https://doi.org/10.1007/JHEP11(2022)036}{\emph{JHEP} {\bfseries 11}
  (2022) 036}, [\href{https://arxiv.org/abs/2208.12660}{{\ttfamily
  2208.12660}}].

\bibitem{Mazzitelli:2023znt}
J.~Mazzitelli, A.~Ratti, M.~Wiesemann and G.~Zanderighi, \emph{{B-hadron
  production at the LHC from bottom-quark pair production at NNLO+PS}},
  \href{https://doi.org/10.1016/j.physletb.2023.137991}{\emph{Phys. Lett. B}
  {\bfseries 843} (2023) 137991},
  [\href{https://arxiv.org/abs/2302.01645}{{\ttfamily 2302.01645}}].

\bibitem{Biello:2024vdh}
C.~Biello, A.~Sankar, M.~Wiesemann and G.~Zanderighi, \emph{{NNLO+PS
  predictions for Higgs production through bottom-quark annihilation with
  MINNLO$_{\rm PS}$}},
  \href{https://doi.org/10.1140/epjc/s10052-024-12845-z}{\emph{Eur. Phys. J. C}
  {\bfseries 84} (2024) 479},
  [\href{https://arxiv.org/abs/2402.04025}{{\ttfamily 2402.04025}}].

\bibitem{Biello:2024pgo}
C.~Biello, J.~Mazzitelli, A.~Sankar, M.~Wiesemann and G.~Zanderighi,
  \emph{{Higgs boson production in association with massive bottom quarks at
  NNLO+PS}}, \href{https://doi.org/10.1007/JHEP04(2025)088}{\emph{JHEP}
  {\bfseries 04} (2025) 088},
  [\href{https://arxiv.org/abs/2412.09510}{{\ttfamily 2412.09510}}].

\bibitem{Biello:2026nhj}
C.~Biello, C.~Savoini, C.~Signorile-Signorile and M.~Wiesemann,
  \emph{{Next-to-next-to-leading order event generation for $t\bar{t}H$
  production with approximate two-loop amplitude}},
  \href{https://arxiv.org/abs/2603.06143}{{\ttfamily 2603.06143}}.

\bibitem{Nason:2004rx}
P.~Nason, \emph{{A New method for combining NLO QCD with shower Monte Carlo
  algorithms}},
  \href{https://doi.org/10.1088/1126-6708/2004/11/040}{\emph{JHEP} {\bfseries
  11} (2004) 040}, [\href{https://arxiv.org/abs/hep-ph/0409146}{{\ttfamily
  hep-ph/0409146}}].

\bibitem{Frixione:2007vw}
S.~Frixione, P.~Nason and C.~Oleari, \emph{{Matching NLO QCD computations with
  Parton Shower simulations: the POWHEG method}},
  \href{https://doi.org/10.1088/1126-6708/2007/11/070}{\emph{JHEP} {\bfseries
  11} (2007) 070}, [\href{https://arxiv.org/abs/0709.2092}{{\ttfamily
  0709.2092}}].

\bibitem{Alioli:2010xd}
S.~Alioli, P.~Nason, C.~Oleari and E.~Re, \emph{{A general framework for
  implementing NLO calculations in shower Monte Carlo programs: the POWHEG
  BOX}}, \href{https://doi.org/10.1007/JHEP06(2010)043}{\emph{JHEP} {\bfseries
  06} (2010) 043}, [\href{https://arxiv.org/abs/1002.2581}{{\ttfamily
  1002.2581}}].

\bibitem{Cascioli:2011va}
F.~Cascioli, P.~Maierhöfer and S.~Pozzorini, \emph{{Scattering Amplitudes with
  Open Loops}},
  \href{https://doi.org/10.1103/PhysRevLett.108.111601}{\emph{Phys. Rev. Lett.}
  {\bfseries 108} (2012) 111601},
  [\href{https://arxiv.org/abs/1111.5206}{{\ttfamily 1111.5206}}].

\bibitem{Buccioni:2017yxi}
F.~Buccioni, S.~Pozzorini and M.~F. Zoller, \emph{{On-the-fly reduction of open
  loops}}, \href{https://doi.org/10.1140/epjc/s10052-018-5562-1}{\emph{Eur.
  Phys. J. C} {\bfseries 78} (2018) 70},
  [\href{https://arxiv.org/abs/1710.11452}{{\ttfamily 1710.11452}}].

\bibitem{Buccioni:2019sur}
F.~Buccioni, J.-N. Lang, J.~M. Lindert, P.~Maierhöfer, S.~Pozzorini, H.~Zhang
  et~al., \emph{{OpenLoops 2}},
  \href{https://doi.org/10.1140/epjc/s10052-019-7306-2}{\emph{Eur. Phys. J. C}
  {\bfseries 79} (2019) 866},
  [\href{https://arxiv.org/abs/1907.13071}{{\ttfamily 1907.13071}}].

\bibitem{Becher:2009qa}
T.~Becher and M.~Neubert, \emph{{On the Structure of Infrared Singularities of
  Gauge-Theory Amplitudes}},
  \href{https://doi.org/10.1088/1126-6708/2009/06/081}{\emph{JHEP} {\bfseries
  06} (2009) 081}, [\href{https://arxiv.org/abs/0903.1126}{{\ttfamily
  0903.1126}}]. [Erratum: JHEP 11, 024 (2013)].

\bibitem{Becher:2010tm}
T.~Becher and M.~Neubert, \emph{{Drell-Yan Production at Small $q_T$,
  Transverse Parton Distributions and the Collinear Anomaly}},
  \href{https://doi.org/10.1140/epjc/s10052-011-1665-7}{\emph{Eur. Phys. J. C}
  {\bfseries 71} (2011) 1665},
  [\href{https://arxiv.org/abs/1007.4005}{{\ttfamily 1007.4005}}].

\bibitem{Heinrich:2017kxx}
G.~Heinrich, S.~P. Jones, M.~Kerner, G.~Luisoni and E.~Vryonidou, \emph{{NLO
  predictions for Higgs boson pair production with full top quark mass
  dependence matched to parton showers}},
  \href{https://doi.org/10.1007/JHEP08(2017)088}{\emph{JHEP} {\bfseries 08}
  (2017) 088}, [\href{https://arxiv.org/abs/1703.09252}{{\ttfamily
  1703.09252}}].

\bibitem{Davies:2019dfy}
J.~Davies, G.~Heinrich, S.~P. Jones, M.~Kerner, G.~Mishima, M.~Steinhauser
  et~al., \emph{{Double Higgs boson production at NLO: combining the exact
  numerical result and high-energy expansion}},
  \href{https://doi.org/10.1007/JHEP11(2019)024}{\emph{JHEP} {\bfseries 11}
  (2019) 024}, [\href{https://arxiv.org/abs/1907.06408}{{\ttfamily
  1907.06408}}].

\bibitem{Davies:2025qjr}
J.~Davies, K.~Sch{\"o}nwald, M.~Steinhauser and D.~Stremmer, \emph{{ggxy: A
  flexible library to compute gluon-induced cross sections}},
  \href{https://doi.org/10.1016/j.cpc.2025.109933}{\emph{Comput. Phys. Commun.}
  {\bfseries 320} (2026) 109933},
  [\href{https://arxiv.org/abs/2506.04323}{{\ttfamily 2506.04323}}].

\bibitem{Sjostrand:2014zea}
T.~Sjöstrand, S.~Ask, J.~R. Christiansen, R.~Corke, N.~Desai, P.~Ilten et~al.,
  \emph{{An Introduction to PYTHIA 8.2}},
  \href{https://doi.org/10.1016/j.cpc.2015.01.024}{\emph{Comput. Phys. Commun.}
  {\bfseries 191} (2015) 159--177},
  [\href{https://arxiv.org/abs/1410.3012}{{\ttfamily 1410.3012}}].

\bibitem{Grazzini:2017mhc}
M.~Grazzini, S.~Kallweit and M.~Wiesemann, \emph{{Fully differential NNLO
  computations with MATRIX}},
  \href{https://doi.org/10.1140/epjc/s10052-018-5771-7}{\emph{Eur. Phys. J. C}
  {\bfseries 78} (2018) 537},
  [\href{https://arxiv.org/abs/1711.06631}{{\ttfamily 1711.06631}}].

\bibitem{ATLAS:2025hhd}
{\scshape ATLAS} collaboration, G.~Aad et~al., \emph{{Study of Higgs boson pair
  production in the $HH \rightarrow b \overline{b} \gamma\gamma$ final state
  with 308 fb$^{-1}$ of data collected at $\sqrt{s} = 13$ TeV and 13.6 TeV by
  the ATLAS experiment}},
  \href{https://doi.org/10.1016/j.physletb.2026.140280}{\emph{Phys. Lett. B}
  {\bfseries 876} (2026) 140280},
  [\href{https://arxiv.org/abs/2507.03495}{{\ttfamily 2507.03495}}].

\bibitem{Cacciari:2008gp}
M.~Cacciari, G.~P. Salam and G.~Soyez, \emph{{The anti-$k_t$ jet clustering
  algorithm}}, \href{https://doi.org/10.1088/1126-6708/2008/04/063}{\emph{JHEP}
  {\bfseries 04} (2008) 063},
  [\href{https://arxiv.org/abs/0802.1189}{{\ttfamily 0802.1189}}].

\bibitem{ATLAS:2023qzf}
{\scshape ATLAS} collaboration, G.~Aad et~al., \emph{{Search for nonresonant
  pair production of Higgs bosons in the $b\overline{b}b\overline{b}$ final
  state in pp collisions at $\sqrt{s} = 13$ TeV with the ATLAS detector}},
  \href{https://doi.org/10.1103/PhysRevD.108.052003}{\emph{Phys. Rev. D}
  {\bfseries 108} (2023) 052003},
  [\href{https://arxiv.org/abs/2301.03212}{{\ttfamily 2301.03212}}].

\bibitem{ATLAS:2025eii}
{\scshape ATLAS, CMS} collaboration, G.~Aad et~al., \emph{{Highlights of the
  HL-LHC physics projections by ATLAS and CMS}},
  \href{https://arxiv.org/abs/2504.00672}{{\ttfamily 2504.00672}}.

\end{thebibliography}\endgroup

\end{document}